\documentclass[12pt,preprint]{aastex}

\slugcomment{Published in the September 2004 issue of PASP}

\shorttitle{An Algorithm for Pointing Refinement}
\shortauthors{Masci, F. J. et al.}

\begin{document}


\title{A Robust Algorithm for the Pointing Refinement and Registration 
       of Astronomical Images}


\author{Frank J. Masci, David Makovoz and Mehrdad Moshir}
\affil{Spitzer Science Center, California Institute of Technology,
       Pasadena, CA 91125}
\email{fmasci@ipac.caltech.edu}



\begin{abstract}

We present a generic algorithm for performing astronomical
image registration and pointing refinement.
The method is based on matching the positions and fluxes of 
available point sources in image overlap regions. This information
is used to compute a set of image offset corrections by globally
minimizing a weighted sum of all matched point-source
positional differences in a pre-specified reference image frame. A fast linear
sparse matrix solver is used for the minimization. From these corrections,
the pointings and orientations of images can be refined in either a
relative sense where pointings become fixed (registered) relative
to a single input image, or, in an absolute sense (in the ICRS)
if absolute point source information is known. The latter provides
absolute pointing refinement to an accuracy depending on the
robustness of point source extractions, match statistics, and accuracy
of the astrometric catalog used. The software is currently used in the
{\it Spitzer} image processing pipelines, although it is adaptable to
any astronomical imaging system which uses the FITS image format and
WCS pointing standard.

We test the algorithm using Monte Carlo simulations and compare them to 
image data acquired with the Infrared Array Camera (IRAC) on board
the {\it Spitzer Space Telescope}. We find that dispersions in matched
source separations after refinement are entirely consistent with
centroiding errors in source extractions, implying that systematic uncertainties
due to inaccurately calibrated distortions are negligible.
For these data, we predict refinements to better than $\sim70$
and $\sim280$ mas ($2\sigma$ radial) for the IRAC $3.6$ and
$8\micron$ bands respectively. These bands
bracket two extremes in available source matches and, for the data
under study, correspond to an average of about 55 and 8
matches per image in these two bands respectively.

\end{abstract}


\section{Introduction}

The need to register or co-align images in astronomy to an accuracy
better than the Nyquist sampling density of a detector's point
response function is pivotal if one wants to maximize the resolution
and signal-to-noise attainable in a mosaic of co-added images.
The absolute accuracy is inhibited by instabilities in telescope
pointing control, tracking sensors, and how these behave in the
science instrument frame in the presence of thermo-mechanical
disturbances. For instance, imaging detectors on both the
{\it Hubble}\footnote{http://www.stsci.edu/hst/observatory/pointing}
and {\it Spitzer}\footnote{http://ssc.spitzer.caltech.edu/documents/SOM/}
Space Telescopes provide an {\it absolute} pointing accuracy of $\approx$ 0.5-1
arcsec (1-$\sigma$ radial). For comparison, the highest attainable
resolutions\footnote{Measured in terms of the Full-Width Half-Maximum
(FWHM) of the center of an Airy disk pattern} are $\approx$ 0.02 and 1 arcsec
for these telescopes resepectively. Without further refinement,
the current pointing accuracies are insufficient to exploit the
near-diffraction limited resolution capabilities the detectors can provide.
Factors of 10-20 improvement in pointing are required for optimal
image registration.  

Good image registration enables extraction and position determination
of sources to fainter flux levels for a given signal-to-noise ratio.
Comparison or registration with astrometric sources whose positions
are known to better than a few percent of the observed image pointing
uncertainty also allows refinement of image frame pointings in the
International Coordinate Reference System (ICRS).
Absolute pointing refinement can be achieved to an accuracy approaching
that of the astrometric catalog used or better given good match statistics.
This alleviates possible ambiguities when
performing cross-identification/correlation of extracted sources across
wavelength dependent catalogs. Furthermore, the accurate placement of
slits for follow-up spectroscopic studies requires source positions
known to better than a few tenths of an arcsecond in the ICRS, or until
the desired positional accuracy is limited by centroiding error
in the array frame.

Broadly speaking, image registration methods can be loosely
divided into three classes:
algorithms which use information in pixel space directly, e.g., by
correlating common objects \citep{BS72}; algorithms
which attempt to match features or identified parts of objects known
as ``graph-theoretic'' methods \citep{B92}; and algorithms which
use the frequency domain e.g., methods based on computing
cross-correlation power spectra via the Fast Fourier Transform
\citep{KH75}. The two conventional methods for registering
images in {\it astronomy} involve either interactively identifying common point
sources in overlapping image fields (i.e., object correlation), or,
using actual detector acquired pointings (with inherent uncertainties) directly 
and estimating relative image offsets therefrom. These methods are available
in most data reduction packages (e.g. IRAF - \citet{D96};
STARLINK - \citet{B02}) and are mostly limited by accuracies in
point source centroids, sufficient match statistics,
telescope pointing control, or, subject to random-walk
(cumulative) uncertainties.

A number of robust methods for astrometric calibration of single images
have also been implemented in the commonly used data reduction packages
\citep{Ver98, Val98, BC03},
although an automated, self-consistent means for simultaneous registration
and refinement of multiple astronomical images comprising a mosaicked region
is generally lacking. The single image methods assume one has
sufficient numbers of astrometric matches to mitigate against
uncertainties in the centroids of extracted sources.
We have gone a step further by using the available point-source
content to obtain a global refined solution for all frames such that
all matched point-source positional discrepancies in all frame-to-frame and
frame-to-absolute overlaps are minimized. Combining the
relative and absolute source information reduces the demand on
having sufficient astrometric matches. This becomes important for mid to
far-infrared imagery where cross-wavelength
astrometric calibration is often unreliable due to differing
sensitivities, source populations and detector point spread function sizes.

We have developed an algorithm to register and refine simultaneously
the pointing of an ensemble of astronomical images
to accuracies better than that inherent in point source centroid
uncertainties (and dictated by point source match statistics).
This paper describes the global minimization algorithm and 
presents a case study using data from {\it Spitzer's} 
Infrared Array Camera (IRAC). The outline is as follows. Section~\ref{algor}
describes the algorithm and pointing refinement accuracies expected therefrom.
Section~\ref{sim} validates the algorithm using a Monte Carlo simulation of
IRAC data. Section~\ref{iracobs} compares these results to real observations 
acquired with IRAC. Discussion  and conclusions are given in
Section~\ref{conc}.

\section{Algorithm}\label{algor}

The algorithm has been implemented into a stand-alone software package
called {\it pointingrefine}. The main inputs to {\it pointingrefine}
are a FITS image list, with each FITS image containing the standard
World Coordinate System (WCS) keywords \citep{GC02,Cal02},
an accompanying list of flux-calibrated point-source extraction tables,
optional astrometric source list, and configuration
parameters. The software expects point source extraction tables
adhering to the format generated by the {\it Spitzer Science Center}
(SSC) source extractor. The complete
{\it pointingrefine} package is available at
{\it http://ssc.spitzer.caltech.edu/postbcd/}. This includes programs
to perform point source extraction as well, although these will
not be described here. A general overview of the processing steps
involved in the global minimization algorithm (the heart of 
{\it pointingrefine}) is shown in Figure~\ref{fig1}. In the following
sections, we expand on some of the more important computational steps
of this algorithm.

\subsection{Set-up and Point Source Matching}\label{setup}

Prior to source matching, we first compute the total number of
image-pairs expected to be overlapping (which could potentially contain
correlated/common sources) in the input ensemble of images. This is
used for efficient a-priori memory allocation.
Given a number of images, $N_{imgs}$, the maximum number of
distict frame pairs that can mutually overlap is

\begin{equation}\label{npairs}
N_{maxpairs} = \frac{1}{2}N_{imgs}\left(N_{imgs} - 1\right).
\end{equation}

\noindent This maximum occurs when all images are stacked
more-or-less on top of each other. For a sparse
mosaic, this number is smaller and thus puts less of a burden
on the required memory. The total number of potential overlapping frame pairs
is found by computing the distances between the centers of every
image pair and finding whether the distance is less than the typical radius
of a circle enscribing an image.

Source positions from all extraction tables (including absolute astrometric
references if specified) are sorted in declination. This preconditioning
speeds up the source matching procedure, converting it from an $O(N^2)$
to an approximate $O(N)$ linear process. Every possible overlapping image
pair is scanned for common point sources in the RA, Dec coordinate system.
Both position and flux matching is performed. The position
match step attempts to find
sources falling within a nominal search radius (typically several times the
root sum squares of prior image pointing uncertainty and extraction
centroid error). If more than one match is
found within the search radius, no match is declared, due to possible
ambiguity. In other words, only singly matched sources
within the search radius are used. Sources are also
simultaneously matched in flux. A flux match is
satisfied if two fluxes fall within a maximum tolerable relative flux
difference threshold. The software includes options to re-scale the input
fluxes of astrometric references to be commensurate with those of
actual extractions.
A minimum of two matches per image is enforced to declare a
correlated image pair, since this is the minimum required to 
estimate a rotational offset between the pair unambiguously.

In preparation for the global minimization step (see below),
all point source match positions and uncertainties are mapped into
the pixel coordinates of a Cartesian reference image frame.
The definition of this reference depends on whether absolute
or relative refinement is desired. The {\it pointingrefine} software
distinguishes between these options from its given inputs. Two inputs
are required for absolute refinement. First, the software expects a
Fiducial Image Frame (FIF). This is a file listing various WCS
parameters that define the tangent point and boundary of a
``fiducial image'' encompassing all images.
Second, a list of astrometric (absolute)
sources that fall within the FIF boundary is required. The FIF
acts as an effective input image whose sources are the astrometric references.
Refinement with respect to such a FIF ensures both absolute and relative
refinement amongst images. In relative refinement mode,
all input images can be registered
and refined with respect to a single input image. In this mode,
either the software selects an image from the input list which is maximally
correlated (has most overlaps) with other images, or the user can specify
their own.

\notetoeditor{Please place Figure~\ref{fig1} here: fig1.eps}

\subsection{Global Minimization}\label{GM}

Consider the simple three-image mosaic in Figure~\ref{fig2}. Image ``1''
defines the ``fiducial'' reference frame. The circles represent point sources
detected from each overlapping image pair transformed into the reference frame
of image 1. The filled circles are sources extracted from image {\it n} and the open
circles are sources extracted from either image 1 or {\it m}. The matched
source pairs are shown offset from each other to mimic the presence of
random pointing uncertainty in each input image. These are the offsets we
wish to compute and use to correct each frame pointing. A one-dimensional
representation of the projection geometry
in the reference image frame is depicted in Figure~\ref{fig3}.
All input images have sizes in the reference frame that
depend on their distance from the reference image tangent point.
The projected linear size scales with angular distance $\theta_t$
as $\:\:\approx 1\:+\:\tan^{2}\theta_t$.
The {\it pointingrefine} algorithm appropriately accounts for the
inflation of centroid uncertainties and separations of correlated sources
when projected into the reference frame. A potential problem for the
algorithm is if image sizes and mosaic extents are large enough to cause a
non-uniform dependence in scale over the region in which an
input image is projected. The dependence of this effect with
$\theta_t$, however, is weak. For instance, the projected scale varies
by $\lesssim1\%$ across an image $30\arcmin$ in size at angular distances
$\theta_t\lesssim30\degr$.

The main assumption of the algorithm is that the random uncertainty
in measured twist angle\footnote{We define the ``twist angle'' as the
relative orientation of an image in an orthogonal coordinate system, {\it not}
the conventional position angle measured in the ICRS}
of an individual image frame ($\delta\theta^m$ in Figure~\ref{fig2})
is small enough to ensure
$\sin\delta\theta^m\approx\delta\theta^m$ (see below).
$\delta\theta\lesssim1\arcmin$ is a good working measure for
the purposes of this algorithm.
This is justified for the {\it Spitzer} science instruments where the
absolute twist angle uncertainty\footnote{From Jet Propulsion Laboratory 
Internal Document: SIRTF Instrument Pointing Frame Kalman Filter
Execution Summary (IPF Team), report ID01M095, October 2003}
is typically $\lesssim30\arcsec$ (1-$\sigma$).

\notetoeditor{Please place Figure~\ref{fig2} here: fig2.eps}

In a rectilinear coordinate system (say that defined by the frame of image 1 in
Figure~\ref{fig2}), the positions of a point source {\it i} detected from
images {\it m} and {\it n} are related via the transformation:

\begin{equation}\label{fulltran}
\left(\begin{array}{c}
x^m_i \\
y^m_i
\end{array}\right)\rightarrow
\left(\begin{array}{c}
\tilde{x}^n_i \\
\tilde{y}^n_i
\end{array}\right)=
\left(\begin{array}{c}
x^m_c \\
y^m_c
\end{array}\right)+
\left(\begin{array}{lr}
\cos\delta\theta^m & -\sin\delta\theta^m\\
\sin\delta\theta^m & \cos\delta\theta^m
\end{array}\right)
\left(\begin{array}{c}
x^m_i - x^m_c \\
y^m_i - y^m_c
\end{array}\right)+
\left(\begin{array}{c}
\delta X^m \\
\delta Y^m
\end{array}\right)
\end{equation}

\noindent where $\delta\theta^m$ is a rotational offset, $\delta X^m$,
$\delta Y^m$ orthogonal translations, and ($x^m_c$ , $y^m_c$) the
center coordinates of image {\it m} in the
reference image frame. The rotation $\delta\theta^m$ is measured
in a counterclockwise sense so that a rotation followed by
orthogonal translations in {\it x} and {\it y} of image {\it m}
will align the sources (open circles) detected therein with those
detected in image {\it n}.

\notetoeditor{Please place Figure~\ref{fig3} here: fig3.eps}

The assumption of small twist angle uncertainty (see above)
implies $\sin\delta\theta^m\approx\delta\theta^m$ and
$\cos\delta\theta^m\approx 1$ and thus, the pair of equations defined
by (\ref{fulltran}) can be linearized in $\delta\theta^m$ as follows.

\begin{eqnarray}\label{xytran}
x^m_i \rightarrow \tilde{x}^n_i & = & x^m_i - (y^m_i - y^m_c)\delta \theta^m + \delta X^m \nonumber\\ & & \\
y^m_i \rightarrow \tilde{y}^n_i & = & y^m_i + (x^m_i - x^m_c)\delta \theta^m + \delta Y^m , \nonumber
\end{eqnarray}

\noindent where ($\tilde{x}^n_i$, $\tilde{y}^n_i$) represents a position,
corrected for unknown image rotation and orthogonal translations.

We define a cost function $L$, representing an inverse-variance weighted
sum of squares of all matched point-source positional differences
from all overlapping image pairs ($m$, $n$):

\begin{equation}\label{cost}
L = \sum_{m,n (n>m)}\sum_{i}\left\{\frac{1}{\Delta x_i^{m,n}}[\tilde{x}^n_i - \tilde{x}^m_i]^2 +
\frac{1}{\Delta y_i^{m,n}}[\tilde{y}^n_i - \tilde{y}^m_i]^2\right\}\:\:+\:\:L_{apriori}\:\:,
\end{equation}
where
\begin{eqnarray}
\Delta x_i^{m,n} & = & \sigma^2(x^m_i) + \sigma^2(x^n_i)\\ \nonumber
\Delta y_i^{m,n} & = & \sigma^2(y^m_i) + \sigma^2(y^n_i)\nonumber
\end{eqnarray}

\noindent and the $\sigma^2$ represent variances in extracted point source
centroids. The summations in Equation~\ref{cost} are over all matched
point-sources from all image overlaps, including sources matched
to any astrometric references present in the FIF if available.
The FIF or absolute reference frame (as defined in Section~\ref{setup})
is treated as an effective input image when source matching is performed.

The additive term $L_{apriori}$ in Equation~\ref{cost}
represents an ``a-priori'' weighting function which makes
use of actual measured pointing uncertainties
of the images. This function is defined as:

\begin{equation}\label{apriori}
L_{apriori}\:\:=\:\:\sum_{m,n (n>m)}\left\{
\frac{(\delta X^m)^2}{\sigma^2_{Xm}}+
\frac{(\delta Y^m)^2}{\sigma^2_{Ym}}+
\frac{(\delta \theta^m)^2}{\sigma^2_{\theta m}}+
\frac{(\delta X^n)^2}{\sigma^2_{Xn}}+
\frac{(\delta Y^n)^2}{\sigma^2_{Yn}}+
\frac{(\delta \theta^n)^2}{\sigma^2_{\theta n}}\right\}\:,
\end{equation}

\noindent where the $\sigma^2_{jm}$, $\sigma^2_{jn}$ ($j\:=\:X,\:Y,\:\theta$)
represent {\it measured} pointing variances (in the ICRS) transformed into
the reference image frame. The purpose of including $L_{apriori}$ is to
avoid over-refining or biasing those images whose inherent measured
pointing uncertainties are already small (within nominal requirements).
In other words, those images whose uncertainties are known to be small
a-priori, will have have a larger contribution to $L_{apriori}$ relative
to the correlated source term (double sum in Equation~\ref{cost}).
Consequently, the solution will be biased towards minimizing $L_{apriori}$
and not the correlated source term which could potentially degrade
expected image offsets (and final refined pointings). In the limit
$L\:\rightarrow\:L_{apriori}$, the global minimum will be closer to
$\delta \theta\approx\delta X\approx\delta Y\approx0$. Vice-versa,
large pointing uncertainties will bias the solution towards the
correlated source term where refinement via point-source matches
is obviously needed.

Equation~\ref{cost} can be rewritten in terms of physical image
offsets $\delta \theta^m$, $\delta X^m$ and $\delta Y^m$ for an
arbitrary image {\it m} via use of Equations~\ref{xytran} and \ref{apriori}:

\begin{eqnarray}\label{bigcost}
\nonumber
L&=&\sum_{m,n (n>m)}\Biggl[\:\sum_{i}\Biggl\{\frac{1}{\Delta x_i^{m,n}}
\left[ x^m_i - (y^m_i - y^m_c)\delta \theta^m + \delta X^m
- x^n_i + (y^n_i - y^n_c)\delta \theta^n - \delta X^n\right]^2\:\:+\\
&\:&\frac{1}{\Delta y_i^{m,n}}\nonumber
\left[y^m_i + (x^m_i - x^m_c)\delta \theta^m + \delta Y^m
- y^n_i - (x^n_i - x^n_c)\delta \theta^n + \delta Y^n\right]^2\Biggr\}\:\:+\\
&\:&\frac{(\delta X^m)^2}{\sigma^2_{Xm}}+
\frac{(\delta Y^m)^2}{\sigma^2_{Ym}}+
\frac{(\delta \theta^m)^2}{\sigma^2_{\theta m}}+
\frac{(\delta X^n)^2}{\sigma^2_{Xn}}+
\frac{(\delta Y^n)^2}{\sigma^2_{Yn}}+
\frac{(\delta \theta^n)^2}{\sigma^2_{\theta n}}\:\Biggr].
\end{eqnarray}

\noindent The cost function defined by
Equation~\ref{bigcost} can be treated as a standard
$\chi^2$ statistic to the extent that point source centroid uncertainties
are independently random and Gaussian, a good approximation in this case.
The probability density function for $L$ about its global
minimum is therefore the $\chi^2$ distribution for $\nu$ degrees of freedom where
\begin{equation}\label{dof}
\nu\:=\:2N_{matches} - 3(N_{imgs} - 1).
\end{equation}
$N_{matches}$ is the total number of point-source matches in all
image overlap regions and $N_{imgs}$ the total number of images containing the
detected matches (including the reference image).
The total number of parameters to solve
is actually $3(N_{imgs} - 1)$ since we have three offsets
($\delta \theta^m$, $\delta X^m$ and $\delta Y^m$) for every image {\it m}
and we exclude the reference image which by definition is constrained to have
$\delta \theta = 0$, $\delta X = 0$ and $\delta Y = 0$.

Our aim is to minimize $L$ with
respect to all image offsets $\delta\theta^m$, $\delta X^m$ and
$\delta Y^m$ for every correlated image {\it m}.
At the global minimum of $L$, partial derivatives with respect to these
three offsets for each image {\it m} are required to vanish:

\begin{equation}\label{partials}
\frac{\partial L}{\partial\:\delta\theta^m} = 0;\:\:\:\:
\frac{\partial L}{\partial\:\delta X^m} = 0;\:\:\:\:
\frac{\partial L}{\partial\:\delta Y^m} = 0.
\end{equation}

\noindent Evaluating these partial derivatives leads to a set of three simultaneous
equations for each image in the ensemble. For $N_{imgs}$, we therefore have
$3(N_{imgs}-1)$ simultaneous equations in $3(N_{imgs}-1)$ unknowns.
Each image {\it m} of an ensemble $m\:=\:1..M$ could be potentially
correlated (contain sources in common) with any other image {\it n}
where $n\:=\:1..N;\:m\:\neq\:n$ and $M\:=N\:=\:N_{imgs}-1$. From the conditions
defined in Equation~\ref{partials}, the simultaneous system of equations used
to solve for the offsets of all images {\it m}, can be represented
as the matrix equation:

\begin{equation}\label{matrixeq}
\left(\begin{array}{ccccccccc}
A^{m=1}_{\theta} & A^{m=1}_{X} & A^{m=1}_{Y} & . & . & . & A^{n=N}_{\theta} & A^{n=N}_{X} & A^{n=N}_{Y} \\
B^{m=1}_{\theta} & B^{m=1}_{X} & 0 & . & . & . & B^{n=N}_{\theta} & B^{n=N}_{X} & 0 \\
C^{m=1}_{\theta} & 0 & C^{m=1}_{Y} & . & . & . & C^{n=N}_{\theta} & 0 & C^{n=N}_{Y} \\
.                & .           & .           & . & . & . &  .               & .           & .   \\
.                & .           & .           & . & . & . &  .               & .           & .   \\
.                & .           & .           & . & . & . &  .               & .           & .   \\ 
A^{n=1}_{\theta} & A^{n=1}_{X} & A^{n=1}_{Y} & . & . & . & A^{m=M}_{\theta} & A^{m=M}_{X} & A^{m=M}_{Y} \\
B^{n=1}_{\theta} & B^{n=1}_{X} & 0 & . & . & . & B^{m=M}_{\theta} & B^{m=M}_{X} & 0 \\
C^{n=1}_{\theta} & 0 & C^{n=1}_{Y} & . & . & . & C^{m=M}_{\theta} & 0 & C^{m=M}_{Y} \\
\end{array}\right)
\left(\begin{array}{c}
\delta\theta^{m=1} \\
\delta X^{m=1} \\
\delta Y^{m=1} \\
. \\
. \\
. \\
\delta\theta^{m=M} \\
\delta X^{m=M} \\
\delta Y^{m=M} \\
\end{array}\right)\:\:=\:\:
\left(\begin{array}{c}
\Psi^{m=1}_A \\
\Psi^{m=1}_B \\
\Psi^{m=1}_C \\
. \\
. \\
. \\
\Psi^{m=M}_A \\
\Psi^{m=M}_B \\
\Psi^{m=M}_C \\
\end{array}\right)
\end{equation}

\noindent Equation~\ref{matrixeq} is of the form ${\bf M.X = \Psi }$. Elements
of the coefficient matrix ${\bf M}$ and the right-hand-side column vector
${\bf \Psi }$ are given in Appendix~\ref{appI}. The solution for the
column vector ${\bf X}$ with
$3(N_{imgs}-1)$ unknowns can be obtained using linear sparse
matrix methods since, depending on the mosaic geometry, a large number of
the matrix elements can be zero. In general, the fraction of zeros
will be $\geqslant\frac{2}{9}$ ($\gtrsim22\%$). The $\approx 22\%$ minimum will
occur when {\it every} image of the ensemble is correlated with every other, such
as in a coadded stack. The level of sparsity in ${\bf M}$ will increase
with non-zero elements along a block diagonal if one desires to tie and refine
images with respect to astrometric absolute references {\it alone}. In
this specialized case, {\it n = the reference image} and all elements with
superscript {\it n} in ${\bf M}$ will be zero. In general, the minimum and
maximum possible {\it number} of zeros in the matrix ${\bf M}$ are

\begin{eqnarray}\label{minmax}
N_{min}& = & 2(N_{imgs}-1)^{2}\:\:{\rm and}\\\nonumber
N_{max}& = & 9(N_{imgs}-1)^{2} - 7(N_{imgs}-1)\nonumber
\end{eqnarray}
respectively.

We use the UMFPACK\footnote{http://www.cise.ufl.edu/research/sparse/umfpack/}
library to solve the matrix equation~\ref{matrixeq}. This is adapted for solving
large nonsymmetric matrix systems. The library includes an iterative
scheme to correct solutions for the inevitable accumulation in round-off error
during the LU-factorization stage.

We also compute the full error-covariance matrix for all image offsets
($\delta\theta^m$, $\delta X^m$, $\delta Y^m$), which is given by
the inverse of the coefficient matrix: ${\bf C = M^{-1}}$ (e.g., \citet{press}).
Variances in each offset are along the diagonal of ${\bf C}$ and
covariances are given by off-diagonal elements. The covariance matrix ${\bf C}$
is computed using the same matrix solver as above by repeatedly solving
for each unknown column ${\bf X_c}$ in ${\bf M^{-1}}$ such that 
${\bf M.X_{c}=I_{c}}$ where ${\bf I_{c}}$ is the corresponding column in
the identity matrix. An analysis of the covariance matrix using real
data is presented in Appendix~\ref{appII}. 

Images which are {\it not} correlated (lack point source matches) with
any others in the input ensemble cannot contribute to the globally
minimized cost function $L$. For these images, the {\it pointingrefine}
software explicitly sets their reference frame offsets to zero and
no refinement of their pointing is possible. As a further detail,
the inclusion of the a-priori weighting function $L_{apriori}$
(Equation~\ref{apriori}) guarantees that the matrix
in Equation~\ref{matrixeq} will be non-singular. The priors
provide at least one measurement per image. If these are omitted
from the cost function (Equation~\ref{cost}), then there are cases
where the matrix could be singular.
This can occur if the input image ensemble
contains clusters of correlated images disjoint from each other with a
non-contiguous/broken path existing between the clusters. This
situation leads to an under-represention of images across the full
simultaneous system of equations and the determinant will be zero.
As indicated in the processing flow of Figure~\ref{fig1}, if this
occurs in {\it absolute refinement mode}, a second pass computation
is attempted and only those frames which contain absolute astrometric
matches are used. No relative frame-to-frame information is used and
images are refined based on their absolute source content alone.
The matrix effectively becomes
block diagonal and a non-zero determinant is guaranteed.
In {\it relative refinement mode}, no attempt is made to perform
registration within each disjoint sub-ensemble. Instead, the software
will abort with a message indicating that disjoint clusters exist. 

Once image offset corrections ($\delta\theta^m$, $\delta X^m$, $\delta Y^m$)
and associated uncertainties have been determined in the reference
image frame, the final step involves refining the celestial pointing and
orientation of each image {\it m}. This is performed by correcting the
pointing centers ($x^m_c$, $y^m_c$) of each image in the reference image
frame via Equation~\ref{xytran}, i.e.,

\begin{eqnarray}\label{correct}
x^m_c(new) & = & x^m_c(old) + \delta X^m \\
y^m_c(new) & = & y^m_c(old) + \delta Y^m \nonumber
\end{eqnarray}

\noindent and then transforming back to the sky to yield refined pointings
in the ICRS. Image orientations are refined in a similar manner. For these,
we correct and transform two fiducial points per image to
determine uniquely the refined position angle. The main outputs of the
{\it pointingrefine} software are additional WCS keywords written
to FITS image headers representing refined pointings and orientations
on the sky (see processing flow in Figure~\ref{fig1}).

\subsection{Optimization and Expectations}\label{expectme}

The accuracy in pointing refinement or registration
can be severely limited by possible systematics.
For example, inaccurately calibrated image scale and/or distortions
known a-priori to be position dependent will bias extraction centroids
and have adverse affects on the separations of bonafide source matches
and hence final globally minimized solutions. A method to test
for possible contamination from systematics involves examining
distributions of matched source separations after refinement and
comparing these with that contributed by (random) centroid uncertainties.
This will be performed using real data in Section~\ref{iracobs}.

The presence of absolute-astrometric point sources are an important
ingredient for refinement and registration in general.
These reduce the potential for a biased random walk in refined pointing
with distance of an image from a fiducial reference. Such effects
are also alleviated by using appropriate prior pointing uncertainty
information in the global minimization cost function (see Equation~\ref{cost}).
Absolute astrometric references provide a set of ``anchor points'' to 
which all extractions will be attracted. If one desires
to perform {\it absolute} refinement with little or no astrometric
references, then sufficient numbers of images are needed to
increase the number of potential matches in frame overlap
regions for a single point source. This is needed in order to approach
a normal distribution about the expected {\it absolute}
source position. In other words, large numbers of
correlated source positions will ensure that the mean
position of a correlated source cluster is close to the ``truth'',
or that which will result after refinement (assuming no systematics
as discussed above). Due to the rarity of cases with numerous
image overlaps providing good normally-distributed sampling, it is
advised to use astrometric references wherever possible.

The accuracy to which we can refine the pointings of an ensemble of
mutually correlated images predominately depends on the number of point source
matches available (both relative and absolute). With randomly distributed
uncertainties in point source centroids, it is expected that the
mean separation between matched source positions is approximately
Gaussian after refinement, by virtue of the central limit theorem.
In this limit, the (radial) uncertainty in refinement\footnote{Approximated 
as the uncertainty in the mean source match separation with source positions
weighted by their inverse variances.} of a single
image will scale as

\begin{equation}\label{approx}
\sigma_r\simeq\sqrt{\frac{\sigma_{ext}^2 + \sigma_{abs}^2}
{\frac{N_{ext}}{2}\left[1+\left(\frac{\sigma_{abs}^2}
{\sigma_{ext}^2}\right)\right] + N_{abs}}},
\end{equation}

\noindent where $N_{ext}$ and $N_{abs}$ represent respectively the number of
frame-to-frame and frame-to-absolute source match pairs in
{\it all} overlap regions associated with the image,
$\sigma_{ext}$ is a typical source extraction centroid uncertainty, and
$\sigma_{abs}$ is an astrometric source position uncertainty.
This approximation assumes that $\sigma_{ext} > 0$ or
$\sigma_{abs} > 0$ when either $N_{ext} > 0$ or $N_{abs} > 0$ respectively.
For no astrometric matches, we set $N_{abs} = 0$ and
$\sigma_{abs} = 0$, and Equation~\ref{approx}
reduces to $\sigma_{ext}\sqrt{2/N_{ext}}$.
We expect to measure source extraction centroids to better than
$\approx0.1$ pixel, ($\approx0.121\arcsec$ for {\it Spitzer's} IRAC focal
plane arrays). If we assume for example astrometric positional
errors of $\approx0.2\arcsec$ (conservatively speaking),
then to refine image pointings to
an accuracy better than $\approx0.1\arcsec$ will
require at least five {\it astrometric} point-source
matches per frame if $N_{ext} = 0$, or less, if $N_{ext} > 0$.
One can see that with more point-source matches, the better the refinement.
This assumes the observational setup allows for sufficient frame-to-frame
overlap to ensure good numbers of relative matches $N_{ext}$. If this is
not the case, one will have to resort to using pure astrometric
(absolute source) matches alone. 

To summarize, corrections for optimal refinement will effectively be given by
the {\it magnitude} of frame pointing uncertainties with errors
approximated by Equation~\ref{approx}. The latter assumes that in the
limit of increasing number of matches, the mean source separation
per image overlap region is approximately normally distributed with an
uncertainty determined exclusively by point source centroid uncertainties. 
Any position-dependent systematic offset between
source matches such as nonuniform pixel scale or inaccurately
calibrated distortion will limit the refinement accuracy
to the size of systematic error involved. 

\section{Validation Against a Monte-Carlo Simulation}\label{sim}

We quantitatively assess the performance of the above algorithm
using a simulation of 1000 mosaicked images, with each image's coordinates
modeled with an uncertainty drawn from a Gaussian distribution.
The simulation is generic in that it represents a
good overall representation of the type of data that could be acquired
with modern optical/near-infrared detectors to moderately faint
magnitudes ($m_{opt}\simeq23$ or $m_{nearIR}\simeq19$).
To facilitate a comparison with real observations
in Section~\ref{iracobs}, we have chosen to model the source count
distribution and detector properties with that expected (and more or less
observed) in the $3.6\micron$ band of {\it Spitzer's} IRAC instrument.
A more detailed description will be given in Section~\ref{iracobs}.

The ``truth'' source flux-density distribution was simulated using the
models of \citet{Xu03}. These assume a high galactic latitude
stellar model, several galaxy luminosity functions depending on
galaxy morphological type, and exploit a large library of
spectral energy distributions. This simulation
was used extensively for predicting {\it Spitzer} source populations
\citep{Lons03} and for validating processing pipelines.

\notetoeditor{Please place Figure~\ref{fig4} here which consists of
fig4a.eps and fig4b.eps side-by-side from left to right.}

The simulation steps are as follows.
\begin{enumerate}
\item{A ``truth'' list of random source flux densities was generated
      using the models of \citet{Xu03} covering an area of ~0.45 square
      degrees down to a flux density of $10\mu$ Jy at $3.6\micron$
      (equivalent to $\simeq 18.6$ magnitudes in the Vega system).}
\item{Truth sources were assigned both random and correlated
      positions within the area to be mosaicked. Galaxies were assumed
      to have a weak correlated component with amplitude (excess above random)
      based on an empirical $K$-band 2-point angular correlation function:
      $w(\theta)\simeq0.001(\theta/{\rm deg})^{-0.8}$. A pixel scale of
      $\approx1.21\arcsec$ (characteristic of the IRAC arrays)
      was used when mapping sources to the pixel frame of the mosaic.}
\item{Truth sources were convolved with a point spread function (PSF)\footnote{
      Made from in-flight IRAC band-1 observations by the IRAC
      Instrument Support Team at the {\it SSC}.
      It has FWHM of $\sim1.66\arcsec$ and $\sim42\%$ central pixel flux.}
      scaled by the appropriate source flux. PSF-convolved truth sources
      were mapped into the mosaic frame with no pixel resampling.}
\item{1000 ($256\times256$ pixel) ``truth image'' frames were
      generated from the full mosaic area using a dither and mapping
      strategy which assumed $\approx60\%$ adjacent image overlap.
      The optimal map geometry for this 1000-image set was
      generated by D. Shupe (2002, {\it SSC}, private communication).
      RA, Dec, and Twist angle information is also attached to each
      image at this stage.}
\item{A new set of 1000 images was generated (our simulated control sample)
      with uncertainties added to image pointings and position angles.
      Pointing offsets (prior uncertainties) were
      modeled as Gaussian random deviates along each independent
      orthogonal image axis. These were drawn from a zero-mean Gaussian
      distribution with $\sigma = 0.85\arcsec$ per axis. This choice
      for $\sigma$ is based on a pre-launch pointing knowledge
      of $\sim1.2\arcsec$ (1-$\sigma$ radial) for {\it Spitzer} in the
      science instrument frame\footnote{Jet Propulsion Laboratory 
      Internal Document: Operational Implications of the 
      Time-dependent Pointing Behavior of SIRTF;
      C. R. Lawrence et al. Version 1.00; August 22, 2000}.
      Even though {\it Spitzer} can now actually
      do better than this by a factor of $\sim1.5$ (Section~\ref{iracobs}),
      the ultimately refined pointings are independent of the magnitude of
      reasonable simulated prior uncertainty assumed (see below).
      Twist angle uncertainties were modeled as Gaussian with
      $\sigma = 20\arcsec$. A smoothly varying background adjusted
      with the expected Poisson and (IRAC band-1) read-noise
      per pixel was added to each image.}
\item{An absolute source list (representing astrometric references)
      was generated by taking the brightest {\it truth} sources which gave a mean
      density of $\simeq50$ sources per $5.2\arcmin\times5.2\arcmin$
      image region. This resulted in 3030 {\it truth} sources. To imitate an
      astrometric catalog, the sources were assigned with positions
      modeled as $truth\pm u.\sigma$ along each axis with $u$ drawn from a
      Gaussian distribution with $\sigma=0.06\arcsec$. This is typical for
      sources in the {\it 2MASS} point source catalog to $K_s\simeq15$
      (see Section~\ref{iracobs} for details regarding the
      {\it 2MASS} catalog).}
\item{The {\it SSC} point source extractor was used on each simulated
      (control) image to extract sources above a threshold of $5\sigma$.
      This resulted in $\approx40$ extractions per frame.}           
\end{enumerate}

The {\it pointingrefine} software was executed on the 1000 image control sample.
A source match radius of $3.5\arcsec$ was used to comfortably accomodate
prior image pointing errors and extraction centroid errors (typically
$0.15\arcsec$ 1-$\sigma$ per axis). Simultaneous flux matching was also
applied between frame-to-frame and frame-to-absolute (astrometric) matches
with maximum flux difference thresholds of 5 and 10\% respectively.
A zoomed-in ($2.8\arcmin\times3.8\arcmin$) section of our
1000-image simulation (with image pointing errors) is shown in the left panel of
Figure~\ref{fig4}. On the right is the same section after pointing
refinement. The increase in resolution is dramatic. There is a factor
of $\sim6$ decrease in mean source match separation leading to
more localized  point source flux distributions and
detectability to fainter levels.
In this test, the surface brightness is increased by factors of
$\sim2-3.5$ for detected sources after refinement. This is as expected given
that the 1-$\sigma$ radial image pointing uncertainty is of the order
the input pixel size, and the instrinsic PSF has FWHM $\simeq1.66\arcsec$. 
 
To get a more quantitative assessment of the performance of
{\it pointingrefine}, we compare the distribution of separations
between image center pointings of ``truth'' and simulated
(control sample) images before and after refinement. This is shown
in the top panel of Figure~\ref{fig5}. Two different runs of
{\it pointingrefine} were performed based on the number of (brightest)
extractions used per image. One gave an
average of $\sim10$ relative (frame-to-frame) and
$\sim20$ absolute-astromeric
source matches per image (the ``10/20-match'' case), and the second
resulted in an average of $\sim2$ relative and
$\sim3$ absolute matches per image (the ``2/3-match'' case).
It should be noted that two matches per image is the absolute
minimum to determine unambiguously two orthogonal shifts and an
orientation per frame (i.e. the number of degrees of
freedom (\#dof) $=2N_{matches}-3=1$).
The dispersion in image separation relative to truth {\it after refinement}
for the ``10/20-match'' case is $\simeq65$ mas\footnote{1 mas$\:\equiv\:$1
milli-arcsecond} (1-$\sigma$ radial). For the ``2/3-match'' case, this is
$\simeq110$ mas. For the given match statistics, these numbers are
more or less consistent with the simple scaling relation
given by Equation~\ref{approx}. Offset distributions along each axis
are shown in the lower panel of Figure~\ref{fig5} where the open circle
represents the 2-$\sigma$ contour for the ``2/3-match'' case.

\notetoeditor{Please place Figure~\ref{fig5} here which consists of
fig5a.eps, 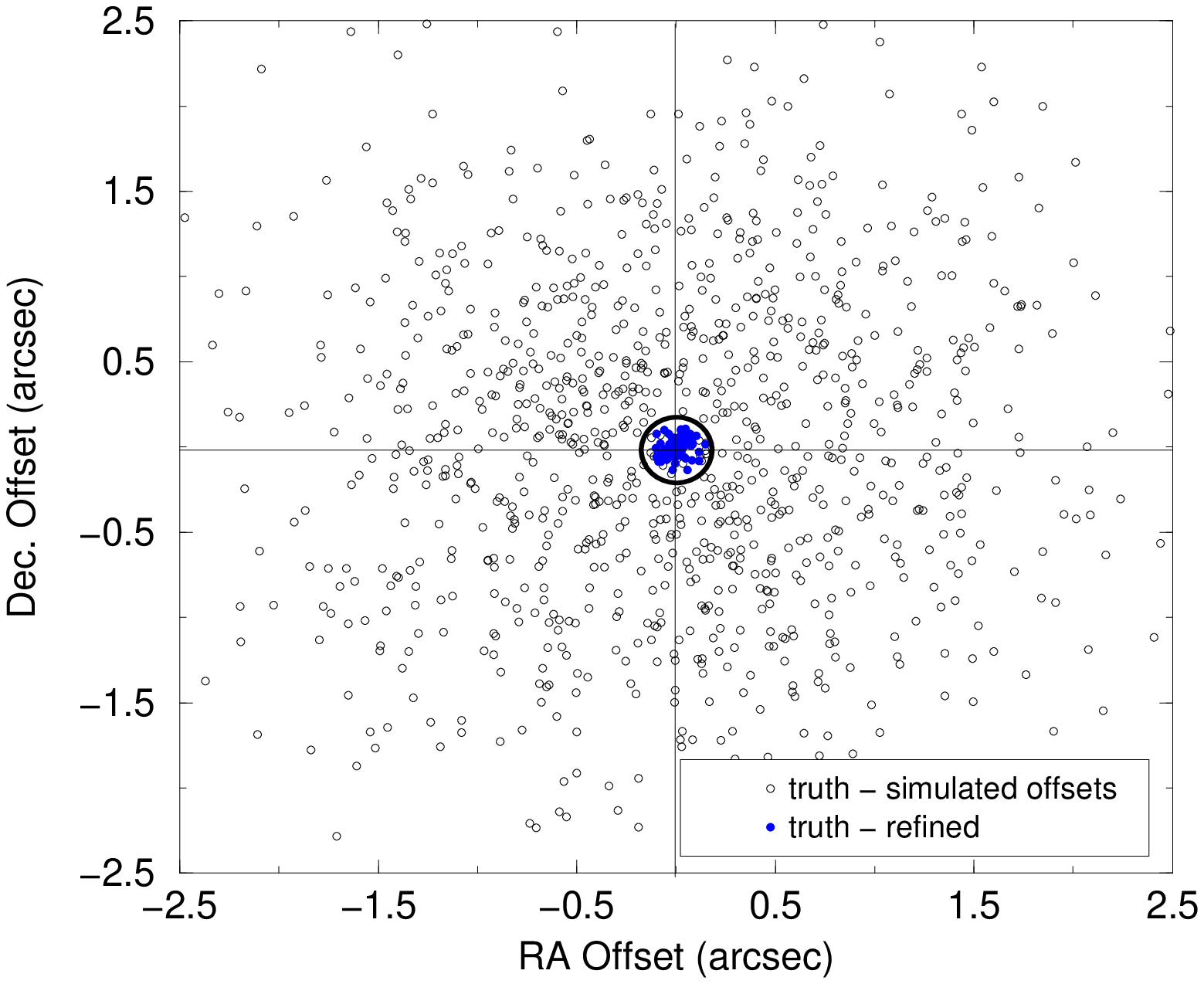 top and bottom respectively.}

By comparing differences in image radial separations before and after
refinement, we find that over 890 images have their pointing refined to
better than 95\% relative to ``truth'' (for the ``10/20-match'' case).
The top panel in Figure~\ref{fig6} shows the distribution
in fractional refinement. This quantity is defined as the ratio of separations:
$1 - {\rm D}(refined - truth)/ {\rm D}(unrefined - refined)$.
For the ``2/3-match'' case, slightly less than half have the
same amount of refinement, although most images are refined to better than
80\%, corresponding to a discrepancy of $\simeq180$ mas
within truth image positions. Thus the refinement is very good, even
with minimal matches.

\notetoeditor{Please place Figure~\ref{fig6} here which consists of
fig6a.eps, 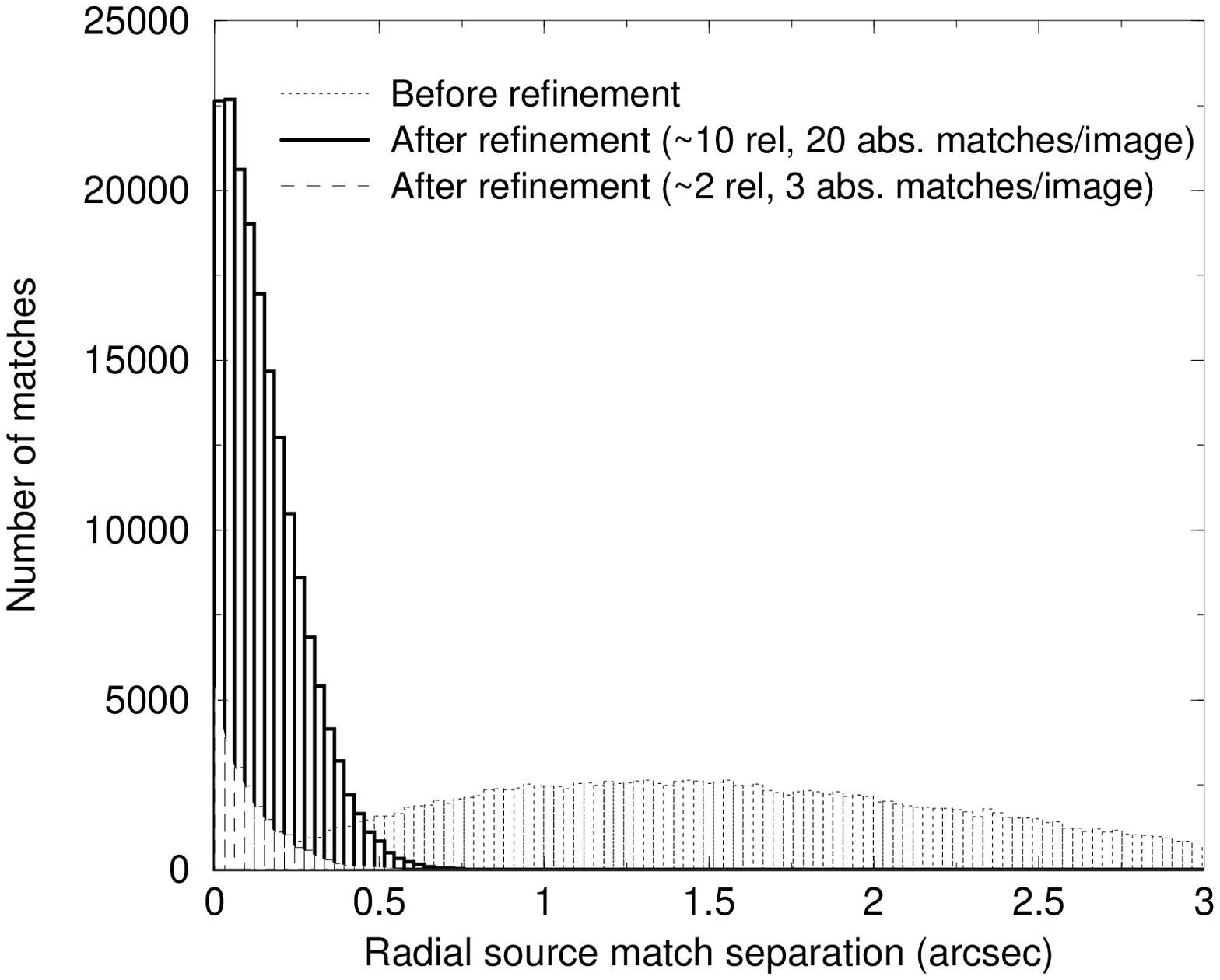 top and bottom respectively.}

The bottom panel in Figure~\ref{fig6} shows distributions in
matched source radial separations before and after refinement.
Uncertainties in both image pointing and source extraction centroids
(or radial separation) contribute to the dispersion in the
{\it unrefined} distribution. Image pointing uncertainties
dominate, with a $1.2\arcsec$ contribution compared to
$\simeq0.2\arcsec$ for extraction centroids (both 1-$\sigma$ radial).
After pointing refinement, the dispersion in radial source separation is
expected to be dominated exclusively by extraction centroid errors
and indeed, the distributions confirm this for both
the ``2/3'' and ``10/20-match'' case (Figure~\ref{fig6} bottom left).
The distribution for the ``2/3-match'' case is narrower since such matches
were performed using the brightest extractions per image, and these
inherently have better determined centroids.

The separation between two source
extraction centroids ($x_i,\:y_i$) and ($x_j,\:y_j$) is given by
\begin{equation}\label{Euclid}
r\:=\:\left[(x_i - x_j)^{2} + (y_i - y_j)^{2}\right]^{1/2}.
\end{equation}
Since a majority of extractions are unresolved point sources
with very circular error-ellipses, we can ignore any correlations
between uncertainties along each axis. To a good approximation,
the uncertainty in radial separation between any two centroids can
therefore be written:
\begin{equation}\label{radunc}
\sigma_{r}\:\simeq\:\sqrt{\sigma_{x_i}^{2} + \sigma_{x_j}^{2}} \:\simeq\:
\sqrt{\sigma_{y_i}^{2} + \sigma_{y_j}^{2}},
\end{equation}

\noindent where ($\sigma_{x_i}^{2},\:\sigma_{y_i}^{2}$) and
($\sigma_{x_j}^{2},\:\sigma_{y_j}^{2}$) are centroid variances
in each axis for sources $i$ and $j$ respectively.
A comparison between uncertainties in matched source radial separation
(Equation~\ref{radunc}) and actual separations (Equation~\ref{Euclid})
after refinement is shown in the top panel of Figure~\ref{fig7}
(for ``10/20-match'' case). They are both mutually consistent,
although the spread is greater at separations $\lesssim0.3\arcsec$.
After refinement and in the absence of systematics, any residual
separation in a matched source pair must be due to intrinsic
centroiding error alone. Since separations
between matched sources along each axis are to a good
approximation independently random and normally distributed with zero mean
($\langle x_i - x_j \rangle\simeq\langle y_i - y_j \rangle\simeq 0$), the
quantity $r$ (Equation~\ref{Euclid}) can be shown to follow a
$\chi$ distribution with two degrees of freedom (e.g., \citet{Evans2000}).
This special case is also known as the Rayleigh distribution:
\begin{equation}\label{Rayleigh}
P(r)\:=\:\frac{r}{\beta^2}
\exp{\left[-\frac{1}{2}\left(\frac{r}{\beta}\right)^{2}\right]},
\end{equation}
where $\beta$ is a parameter characterizing the width.
This can be written in terms of the second moment (variance) of $P(r)$
as follows
\begin{equation}\label{beta2}
\beta\:=\:\sqrt{\left(\frac{2}{4-\pi}\right)\sigma_{r}^{2}}
\:\simeq\:1.52\:\sigma_{r}\:.
\end{equation}

\noindent For a given uncertainty $\sigma_{r}$
as computed from Equation~\ref{radunc},
the variation in the density of points with $r$ along any horizontal
cut in Figure~\ref{fig7} is qualitatively consistent with that
predicted by Equations~\ref{Rayleigh} and~\ref{beta2}.

The bottom panel of Figure~\ref{fig7} shows the dependence of the
reduced $\chi^2$ (effectively the cost function in
Equation~\ref{bigcost} divided by number of degrees of freedom
defined by Equation~\ref{dof}.) as a function of increasing number
of images $N_i$ in our 1000-image simulation. It's important to
note that image offsets are not re-computed using repeated global
minimizations for each new set of images $N_i$. Instead, the original
full 1000-image solution of image offsets is used throughout
to re-compute $\chi^2$ from Equation~\ref{bigcost} as
$N_i$ is increased for all image pairs ($m,\:n$) such that $n<m\leq N_i$.
As one approaches the full image set of $N_i=1000$, one expects
the reduced $\chi^2$ to converge to unity if, on average,
residuals in source separations after refinement are purely
consistent with extraction centroid uncertainties.
The lower reduced $\chi^2$ values for smaller image numbers
(and particularly for all image numbers in the ``2/3-match'' case) 
is due to the nonlinear behavior in $\chi^2$ when the
number of degrees of freedom is small. Better
fits (smaller $\chi^2$) can be obtained for very low numbers of degrees of
freedom. In Figure~\ref{fig7} (bottom panel), only the
``10/20-match'' case (with absolute refinement as studied above)
shows approximate covergence to one, and the other curves are not far from it.
Also shown is a case where only relative
frame-to-frame matches and no astrometric references are used.
This case tends to show a slightly higher reduced $\chi^2$ ($\simeq1.04$),
which is significant since it is almost 20 standard deviations
from the expected value in reduced $\chi^2$ ($\sigma_{\chi^2}=\sqrt{2/{\rm\#dof}}$,
where \#dof$=308935$). This was traced as being due to slightly
under-estimated undertainties in extraction centroids. This is
not seen in the absolute refinement case (solid curve)
since absolute astrometric uncertainties are themselves over-estimated
and their (almost equal) contribution tends to lower the effective $\chi^2$
when combined with relative frame-to-frame matches. 

\notetoeditor{Please place Figure~\ref{fig7} here which consists of
fig7a.eps, 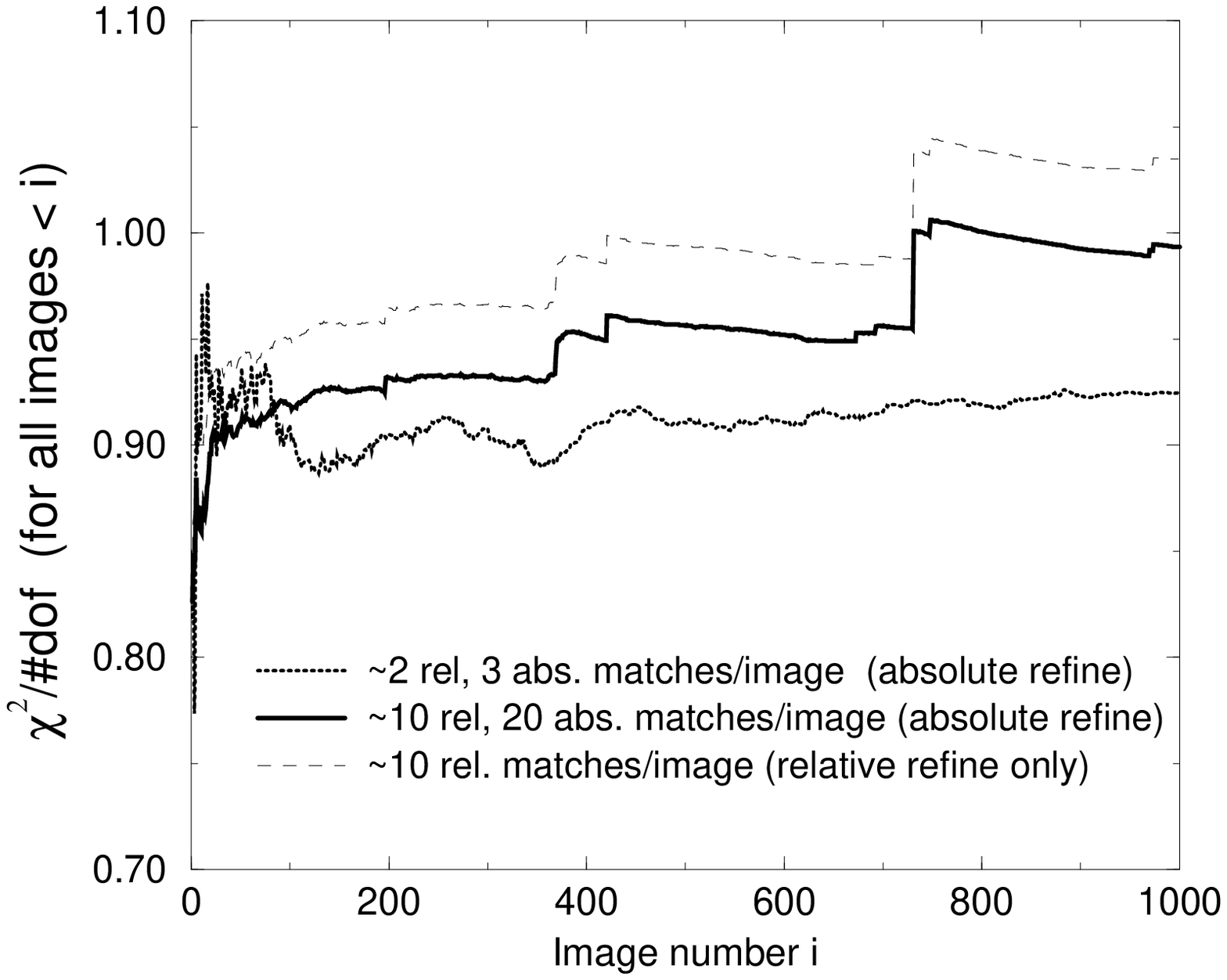, top and bottom respectively.}

To summarize, we have presented a simulation to ascertain the performance
of the {\it pointingrefine} algorithm.
The model-dependent parameters entering our simulation can be isolated to
properties of $3.6\micron$ source populations, specifics of the
IRAC band-1 array, such as PSF and pixelization, and a-priori
telescope pointing knowledge. These can be appropriately rescaled to model
other wavelengths and detectors. However, in the absence of systematics in the
locations of potential matches
between frames (both absolute and relative), and regardless of
instrumental setup or detecter properties, the accuracy
in refined pointing is purely dictated by the accuracy of point
source centroids and match statistics. Our simulation
(Figures~\ref{fig5} and~\ref{fig6}) indeed shows that the refined pointing
will typically never be worse off than the (combined) centroid
uncertainties of matched pairs of sources.
A well sampled and characterized PSF is expected to give centroiding
accuracies to better than one-tenth of a resolution element.
If errors are independently random, good match statistics can then
only work in our favor to give the desired $\sim1/\sqrt{N_{matches}}$ 
improvement in pointing accuracy.

\section{Testing on {\it Spitzer}-IRAC Data}\label{iracobs}

The Infrared Array Camera (IRAC) is one of three focal plane instruments
on the {\it Spitzer Space Telescope} \citep{Fa04}. IRAC provides
simultaneous $\sim5.2\arcmin\times5.2\arcmin$ images at
3.6, 4.5, 5.8 and $8\micron$ (bands 1-4). All four detector arrays in the camera
are $256\times 256$ pixels in size, with a pixel size $\simeq1.2\arcsec$. 
We present here the results of a case study of observations acquired
with IRAC during the in-orbit checkout period (Oct. 2003). In this section,
we validate the pointing performance of IRAC and estimate
the accuracy of refinement that can be achieved using a standard astrometric
catalog and comparisons with our simulation of the $3.6\micron$
band from Section~\ref{sim}. 

\notetoeditor{Please place Figure~\ref{fig8} here: fig8.eps}

The observational request used for our case study consists of 105
regularly spaced images arranged in a rectangular raster pattern. The coverage
map and geometry is shown in Figure~\ref{fig8}. Adjacent images have
$\sim 20\%$ overlap in each axis, with a coverage of 6 and 12
pixels at the edges and (inner) corners respectively.
The mapping was performed in repetitive horizontal scans as
shown by arrows in Figure~\ref{fig8}. The first image is
at top right and the last at bottom left. All images across all bands were
preprocessed for removal of instrumental signatures using the {\it SSC's} IRAC
pipeline\footnote{http://ssc.spitzer.caltech.edu/documents/SOM/} 
and raw pointing information was attached to FITS
headers. Source extraction was then performed using
the {\it SSC} source extractor with PSFs characterized from in-flight data.
Sources were extracted to a uniform signal-to-noise ratio of $5\sigma$
in each band, resulting in an average of $\sim39$, 24, 13 and 6 extractions
per image for bands 1, 2, 3 and 4 respectively. Errors in flux-weighted
centroids were on average $\sim0.18$, 0.22, 0.26 and $0.27\arcsec$
(1-$\sigma$ radial) for each band respectively. 

We used data from the Two-Micron All Sky Survey\footnote{See
http://www.ipac.caltech.edu/2mass/} ({\it 2MASS}) to
define a standard astrometric catalog for all IRAC bands. 
\citet{Zac03} used the USNO CCD Astrograph Catalog (UCAC) (which is accurate
to $\sim20$ mas) to perform an  assessment of the
accuracy of {\it 2MASS} astrometry.
They found random errors of {\it 2MASS} positions of $\sim85$ to 140 mas radial
to a limiting $K_s$ magnitude of 15. We use astrometric sources detected in
the {\it 2MASS} $K_s$ band ($2.12\micron$) which is well suited to IRAC,
and as a good working measure, we impose a magnitude limit of $K_s=15$.
To this limit and at the highest galactic latitudes ($|b|>60\degr$), one
expects to find at least two astrometric
sources per $\sim5.2\arcmin\times5.2\arcmin$ IRAC field $\sim60\%$ of the time.
In the galactic plane anti-center, this increases to at least 30 sources 80\%
of the time (J. Surace 2000, {\it SSC}, private communication).

Using source extractions and astrometric references, the {\it pointingrefine}
software was executed on each of the four, band-dependent 105-image ensembles.
Guided by a post-facto analysis of typical source separations and flux
differences, point-source matching was performed using a nominal search
radius of $2\arcsec$ and simultaneous flux matching performed with
relative (frame-to-frame) and astrometric (frame-to-absolute)
flux difference thresholds of 4\% and 50\% respectively across all bands.
Flux matching thresholds were set conservatively high to account
for relative and absolute photometric calibration errors and instrinsic
scatter between source populations detected in the {\it 2MASS} and IRAC bands.
Relative and astrometric match statistics are summarized in Table~\ref{tab1}.

\notetoeditor{Please place Figure~\ref{fig9} here which consists of
fig9a.eps, 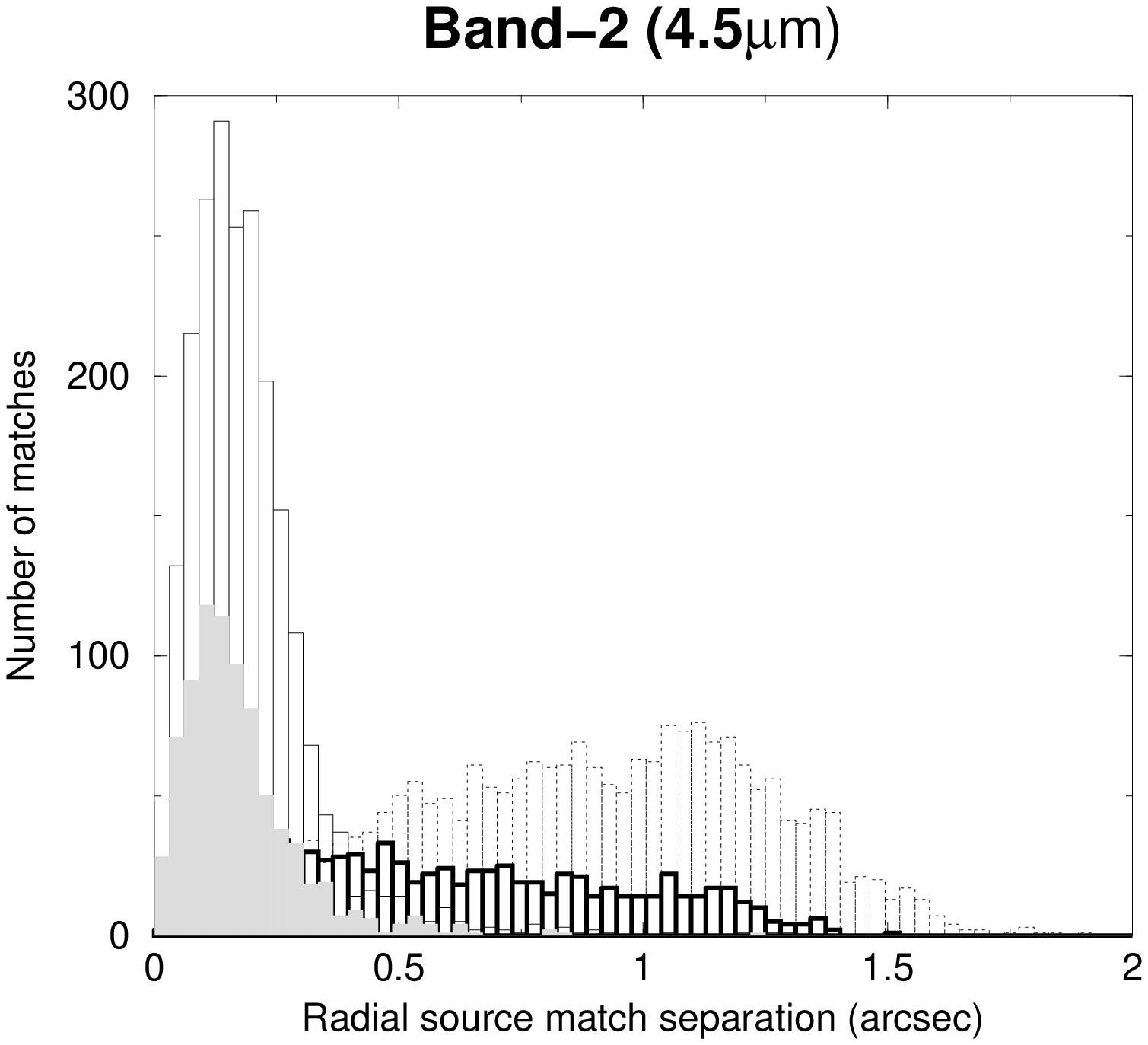, 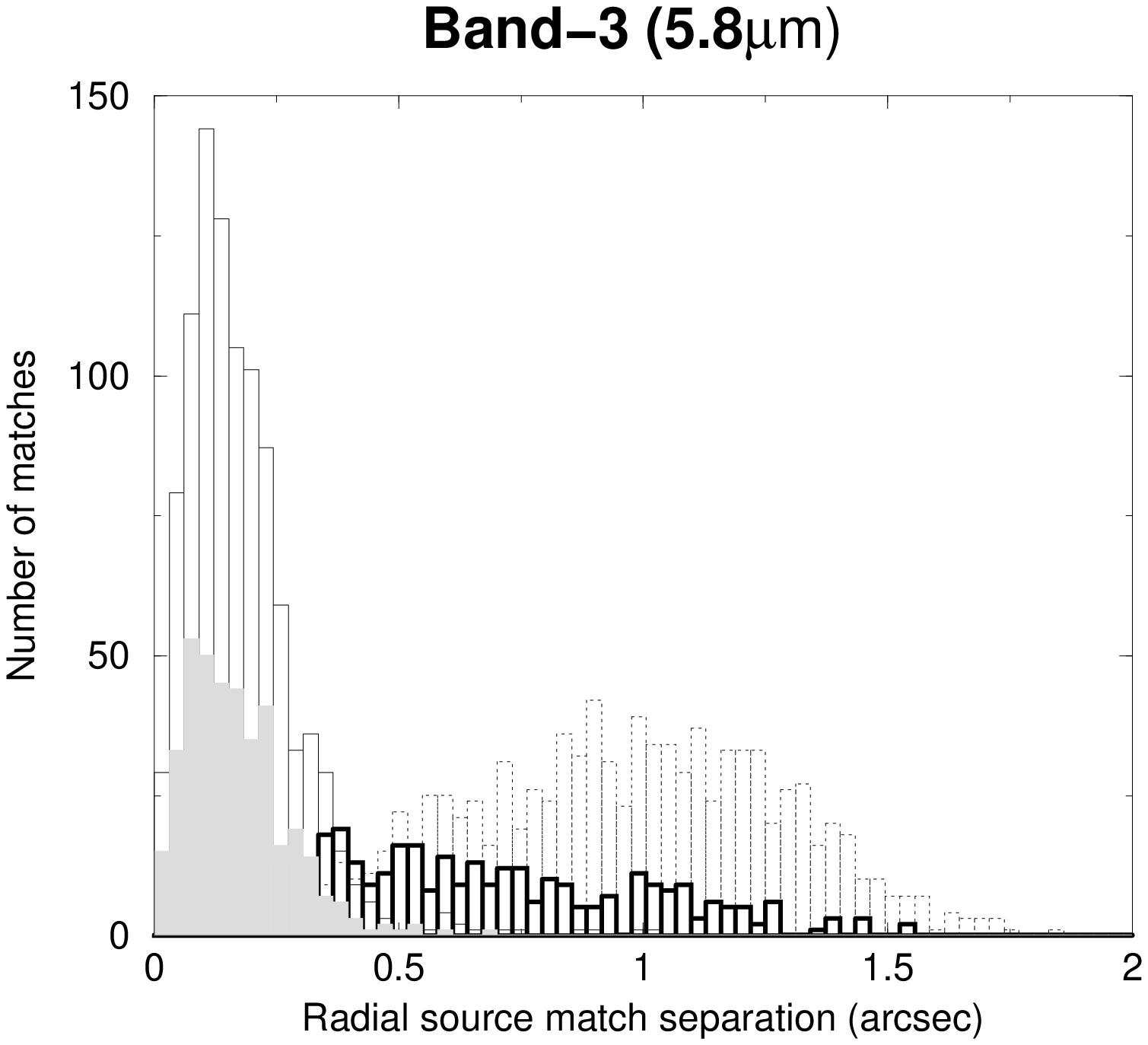 and fig9d.eps arranged in
a 2x2 panel: left,right on top then left,right again under this.}

Figure~\ref{fig9} shows distributions in matched source radial separations
before and after refinement for all bands. Relative and astrometric
matches have been separated. These distributions provide a powerful
diagnostic with which to assess the in-flight pointing
performance in the IRAC science instrument frame. The end-to-end pointing 
accuracy is a function of the inherent star-tracker accuracy, the
spacecraft control system, how well the star-tracker bore-sight is
known in the telescope pointing frame and focal plane
array (science instrument) frame, and variations in these due to
thermo-mechanical deflections. {\it Spitzer's} star-tracker assembly
alone provides pointing and control to better than $0.3\arcsec$ absolute
accuracy. The In-Orbit-Checkout (IOC) dataset analyzed here may
have additional error sources because the images were obtained
while the {\it Spitzer} pointing and data-acquisition systems were
still being tuned and because the image integration times were
very short (1.2 seconds). For example, if the estimate
of the time needed for the spacecraft pointing system to ``settle'' were
slightly too short (i.e., underestimated), then the images would
have been taken while the spacecraft pointing was changing slightly,
providing an additional error term in the pointing transfer.
More standard images with relatively long integration times
would be much less affected by this source (as long as
the settle time is small relative to the integration time).
Therefore, the pointing accuracy we derive for this dataset
may not be typical for {\it Spitzer}, but it still provides a
valid test of the POINTINGREFINE algorithm.

\notetoeditor{Please place Table~\ref{tab1} here}

We can get an estimate of both
the relative and absolute raw pointing for this IRAC dataset
from the ``before refinement''
(dotted and thick lined) histograms in Figure~\ref{fig9}.
The main difference between bands is in
the number of point source matches. Other than that, ranges in radial distributions
are more or less consistent. Across all bands, the 1-$\sigma$ radial separation
between frame-to-absolute matches is $\sim0.85$ to $0.93\arcsec$ ,
while for relative (frame-to-frame) matches, this is $\sim0.61$
to $0.65\arcsec$ (See Table~\ref{tab1}). In addition to
actual pointing dispersion, these estimates include a dispersion
component from point source centroid errors. The contribution from centroiding
error to the ``before refinement'' distributions however is negligible.
Taking for instance the band-1 average centroiding error
of $\sim0.18\arcsec$ and $\sim0.085\arcsec$ for {\it 2MASS}
(1-$\sigma$ radial), this translates to effective match separation
uncertainties of $\sim0.14$ and $\sim0.18\arcsec$ for frame-to-absolute and
frame-to-frame (relative) matches respectively, assuming errors
in each axis are {\it uncorrelated} (see Equation~\ref{radunc}).
This implies that the actual absolute and relative
pointing for the frames in this observation
is typically $\sqrt{0.85^2-0.14^2}\approx0.83\arcsec$
and $\sqrt{0.61^2-0.18^2}\approx0.58\arcsec$ (1-$\sigma$ radial) respectively.
Similar results are found using other bands. We should emphasize
that these estimates are strictly valid for this dataset alone.
They are not representative of the actual pointing performance of IRAC
where the absolute pointing is typically $\sim0\arcsec.45$-$0\arcsec.66$
(1-$\sigma$ radial). For an in-depth study, see \citet{BML04}.

The source separation distributions {\it after} refinement
in Figure~\ref{fig9} allow us to validate how well (or whether) images
were refined to within accuracies determined by source extraction centroids.
As found in our simulation (Figure~\ref{fig5}), images were
refined to better than $\simeq65$ mas (1-$\sigma$ radial) within truth
pointings, with a corresponding $\simeq200$ mas dispersion in
source separations after refinement (Figure~\ref{fig6}).
The top panel in Figure~\ref{fig10} shows the distribution of uncertainties
in radial match separation (from Equation~\ref{radunc}) as a function of
separation $r$ for band-1 {\it after refinement}. These are broadly consistent
above the minimum cut-off uncertainty of $\sim100$ mas imposed 
by the finite size of the PSF. The distribution in $r$ at any given
uncertainty cut $\sigma_{r}$ can also be described by a Rayleigh distribution
(see Equations~\ref{Rayleigh} and~\ref{beta2}). 
Furthermore, Figure~\ref{fig10} shows that typical
systematic uncertainties due to inaccurately calibrated distortion
are minimal, and if present, are expected to be much less
than the centroiding errors.
Guided by the simulation in Section~\ref{sim} and Figure~\ref{fig9},
we conservatively conclude that the majority of image pointings must be refined
to better $\simeq200$ mas. In fact,
we can predict the absolute dispersion in image pointings about truth using
Equation~\ref{approx}. Given the typical match statistics listed in
Table~\ref{tab1} and centroiding errors of $0.18\arcsec$ and $0.27\arcsec$
for bands 1 and 4 respectively, we expect dispersions of
$\sim35$ and $\sim140$ mas (1-$\sigma$ radial) about
truth for these bands respectively. We expect these dispersions to
be smaller once centroiding errors are brought down using
better characterized PSFs.

\notetoeditor{Please place Figure~\ref{fig10} here which consists of
fig10a.eps, 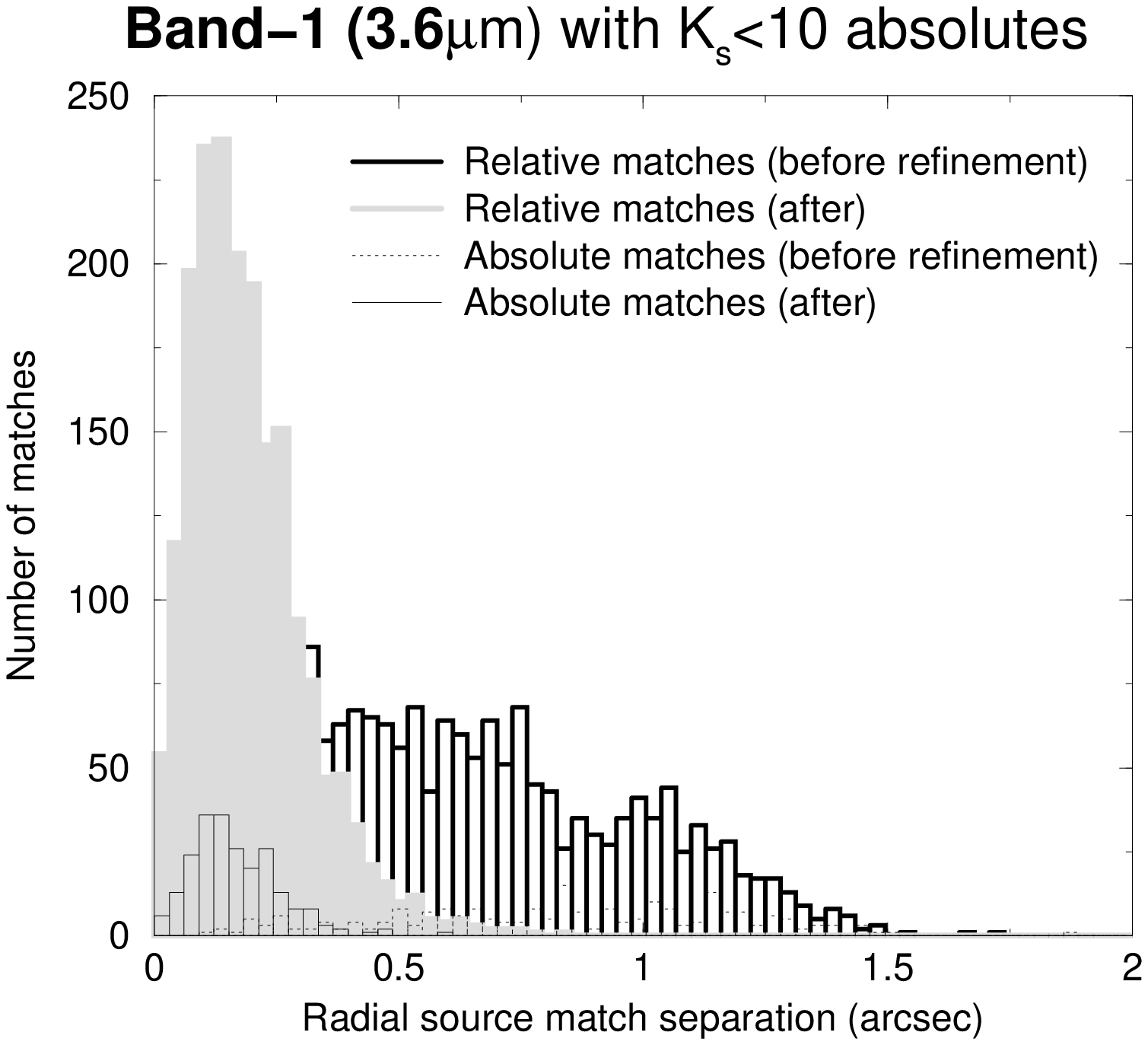, top and bottom respectively.}

In this case study, the number of astrometric (frame-to-absolute) matches are
factors 1.5 to 2 times greater than relative (frame-to-frame) matches
(see Table~\ref{tab1}). This is due to the specific flux difference
thresholds used in source matching (4\% and 50\% for relative and absolute
matches respectively; see above). The results of Figure~\ref{fig9} assumed
{\it 2MASS} point sources with $K_s\leq15$. To ascertain the degree to
which the number of absolute matches control the level of refinement,  
we repeated the source matching and refinement for band-1 using a
{\it 2MASS} magnitude cut of $K_s=10$. The bottom panel of
Figure~\ref{fig10} shows source separations before and after refinement.
The number of absolutes per frame was randomly distributed between
0 and 4 (compared to $\sim20$ relative matches/frame).
When frame-to-absolute separations after refinement are considered alone
(open histogram at bottom left), the mean separation is smaller
by a factor of $\sim2$ compared to the $K_s\leq15$ case
(top left panel of Figure~\ref{fig9}).
This is due to inherently lower positional uncertainties for the
brighter ($K_s\leq10$) sources. When combined with relative matches
however, the mean source separation and hence level
of refinement is essentially unchanged compared to our results above.

\section{Discussion and Conclusions}\label{conc}

We have presented a generic algorithm to perform
astronomical image registration and pointing refinement. It is generic
in the sense that it can be used on any set of astronomical images
which recognize the FITS and WCS pointing standards \citep{GC02, Cal02}.
Either relative (self-consistent frame-to-frame registration),
single image absolute-astrometric, or simultaneous
(relative and absolute) refinement is supported.
The crux of the method involves matching point source positions
between overlapping image frames and using this information
to compute image offset corrections by globally
minimizing a weighted sum of matched point-source positional differences.

To ensure robust registration and refinement, the algorithm is best
optimized with the following criteria.
\begin{enumerate}
\item{The random uncertainty in measured twist angle of an individual image
      frame ($\delta\theta$) is assumed to be small such as to ensure
      $\sin{\delta\theta}\approx\delta\theta$. $\delta\theta\lesssim60\arcmin$
      is a good working measure for the intended applications of this algorithm
      (where $1-\frac{\sin{\delta\theta}}{\delta\theta}\lesssim10^{-4}$).}
\item{Input images have been accurately calibrated for distortion
      and possible nonuniform pixel scale. Any 
      position-dependent systematic offset between source matches
      will limit the refinement accuracy to the size of
      the systematic error involved.}
\item{Sufficient area overlap between adjacent image frames is needed
      to ensure good match statistics.}
\item{Availablity of {\it point} sources with well defined flux
      distribution profiles approaching that of the
      instrument/detector's PSF. Extended sources will lead to
      larger centroiding errors.}
\item{Well characterized PSF(s) for the image(s) at hand. These are crucial
      for accurate determination of source extraction centroids.
      If the inherent telescope pointing uncertainty is of order a third or larger
      than the detector pixel size, centroiding accuracies to better than
      one-tenth of a pixel or resolution element are recommended. For pointing
      uncertainties much less than the pixel size, there is little to be gained
      in resolution by improving the registration.}\label{step}
\item{With the suggested centroiding accuracy from step~\ref{step},
      at least five relative (frame-to-frame) and five absolute source matches
      per frame will give pointings refined to better than three-hundreths of
      a pixel (rms). The greater the number of matches, the better
      the refinement. A minimum of two source matches per image frame, either
      relative, absolute, or both is required to determine all
      offset parameters per image unambiguously.}
\item{Astrometric (absolute) reference sources should be used wherever
      possible. These provide a baseline to counteract any
      systematic deviations from ``truth'', or expected pointing
      in the ICRS, especially if absolute refinement is desired.}
\item{It is assumed that uncertainties in image pointing and point source
      extraction centroids are random and independent.
      Prior image pointing uncertainties should be used if available.
      These will prevent from erroneously over-correcting the pointing
      for cases in which it is well known a-priori.}
\end{enumerate}

Our simulations show that potentially good refinement
can be obtained with minimal requirements. For a large fraction
of images in an ensemble, refinements of better than $\simeq65$ mas
(1-$\sigma$ radial from truth) can be obtained with an average
of $\sim10$ relative and $\sim20$ absolute matches per frame with
extraction centroids $\lesssim0.2\arcsec$. This amounts to an improvement
of 95\% relative to truth for a majority of images. This could be
better with higher match statistics and/or more accurate
centroids, since typically the 1-$\sigma$
dispersion about truth scales as $\sim\sigma_{ext}\sqrt{2/N_{ext}}$ for
a given centroiding uncertainty $\sigma_{ext}$ and number of
matches $N_{ext}$.

Analysis of observations from {\it Spitzer's} IRAC
instrument shows that the dispersion in source separations after
refinement is entirely consistent with the inherent
dispersion in extraction centroid uncertainties. This implies that 
systematic uncertainties such as inaccurately calibrated
distortions are negligible, since otherwise dispersions
in matched source separations after refinement would be larger
relative to centroiding errors.
Comparing dispersions of refined pointings about {\it truth}
with matched source separations as found in our simulation,
and rescaling to the appropriate numbers of matches using
$1/\sqrt{N_{match}}$ scaling, we predict (at the $2\sigma$ level)
refinements to better than $\sim70$ mas and $\sim280$ mas for
IRAC bands 1 and 4 respectively. These bands bracket two
extremes in available source matches, and these refinement estimates
correspond to $\sim55$ and $\sim8$ (astrometric and relative) matches
per band-dependent frame respectively.

The goal of astronomical image registration is to exploit
the resolution capabilities of existing and
upcoming detectors whose pointing control and stability may not evolve
at the same rate. The aim is to optimize the achievable signal-to-noise and
science return therein. The algorithm presented herein is just
the tip of the iceberg at exploring one of many optimization techniques
used in the diverse fields of image and signal processing and
computer vision science. 

\acknowledgments

FJM is indebted to John Fowler and John Stauffer for
proofreading the manuscript and providing useful suggestions.
We thank Howard McCallon for illuminating
discussions and David Shupe for assistance with simulations.
This work is based in part on archival data obtained with the
{\it Spitzer Space Telescope}, which is operated by the Jet Propulsion
Laboratory, California Institute of Technology under NASA contract 1407.
Support for this work was provided by NASA through an award issued
by JPL/Caltech. This publication makes use of data products from
the Two Micron All Sky Survey, which is a joint project of the
University of Massachusetts and the Infrared Processing and
Analysis Center/California Institute of Technology, funded by the
National Aeronautics and Space Administration and the
National Science Foundation.

\appendix

\section{Elements of the Coefficient Matrix}\label{appI}

In this section, we provide general analytic expressions for
elements of the coefficient matrix ${\bf M}$
(Equation~\ref{matrixeq}). These elements are obtained by
applying the minimization conditions (Equation~\ref{partials})
to the cost function defined by Equation~\ref{bigcost}.
The ``base'' coefficent labels $A$, $B$ and $C$ correspond to
the three equations obtained by evaluating the partial derivatives
in Equation~\ref{partials} (labelled equations
$A$, $B$ and $C$ from left to right respectively).
Position and uncertainty variables appearing in the expressions
below were defined in Section~\ref{GM}.

\begin{eqnarray}
A^m_\theta &=&\sum_n\sum_i\frac{(y^m_i - y^m_c)^2}{\Delta x_i^{m,n}} +
              \frac{(x^m_i - x^m_c)^2}{\Delta y_i^{m,n}} +
              \sum_{n}\frac{1}{\sigma^2_{\theta m}} \\ 
A^m_X&=&-\sum_n\sum_i\frac{(y^m_i - y^m_c)}{\Delta x_i^{m,n}}\\
A^m_Y&=&\sum_n\sum_i\frac{(x^m_i - x^m_c)}{\Delta y_i^{m,n}}\\
A^n_\theta &=&-\sum_n\sum_i\frac{(y^m_i - y^m_c)(y^n_i - y^n_c)}{\Delta x_i^{m,n}} +
              \frac{(x^m_i - x^m_c)(x^n_i - x^n_c)}{\Delta y_i^{m,n}}\\
A^n_X&=&\sum_i\frac{(y^m_i - y^m_c)}{\Delta x_i^{m,n}}\\
A^n_Y&=&-\sum_i\frac{(x^m_i - x^m_c)}{\Delta y_i^{m,n}}\\
B^m_\theta &=&A^m_X\\
B^m_X&=&\sum_n\sum_i\frac{1}{\Delta x_i^{m,n}} + \sum_{n}\frac{1}{\sigma^2_{Xm}}\\
B^n_\theta &=&\sum_i\frac{(y^m_i - y^m_c)}{\Delta x_i^{m,n}}\\
B^n_X&=&-\sum_i\frac{1}{\Delta x_i^{m,n}}\\
C^m_\theta &=&A^m_Y\\
C^m_Y&=&\sum_n\sum_i\frac{1}{\Delta y_i^{m,n}} + \sum_{n}\frac{1}{\sigma^2_{Ym}}\\
C^n_\theta &=&-\sum_i\frac{(x^n_i - x^n_c)}{\Delta y_i^{m,n}}\\
C^n_Y&=&-\sum_i\frac{1}{\Delta y_i^{m,n}}\\
\Psi^m_A&=&-\sum_n\sum_i\frac{(y^m_i - y^m_c)(x^n_i - x^m_i)}{\Delta x_i^{m,n}} +
           \frac{(x^m_i - x^m_c)(y^m_i - y^n_i)}{\Delta y_i^{m,n}}\\ 
\Psi^m_B&=&-\sum_n\sum_i\frac{(x^m_i - x^n_i)}{\Delta x_i^{m,n}}\\
\Psi^m_C&=&-\sum_n\sum_i\frac{(y^m_i - y^n_i)}{\Delta y_i^{m,n}}\\\nonumber
\end{eqnarray}

\section{The Error-Covariance Matrix}\label{appII}

The full error-covariance matrix is one of the by-products of the
{\it pointingrefine} software. This reports all variances and
covariances for and between all (inter and intra) image offsets
necessary for refinement from the global minimization. 
It can be used to explore the
strength of long-distance correlations between images in a mosaic
and the presence of any undue systematic walks after refinement.
The latter could arise if one lacks the desired number of absolute
astrometric sources, or correct magnitude of prior image pointing uncertainties
as was discussed in Section~\ref{expectme}.

For a given pair of images ($i,\:j$) and three computed offsets per image
($\delta\theta,\:\delta X,\:\delta Y$), we have a total of six possible
(correlated) offset pairs or covariance matrices. The possible
covariances (or variances if $i=j$ for the same offsets)
for any two images are therefore
\begin{equation}\label{allcovs}
cov(\theta_i,\theta_j),\:\:
cov(\theta_i,X_j),\:\:
cov(\theta_i,Y_j),\:\:
cov(X_i,X_j),\:\:
cov(X_i,Y_j),\:\:
cov(Y_i,Y_j)
\end{equation}
If we define any of these covariances generically as
$cov(\alpha_i,\:\beta_j)$ then the format for the error-covariance
matrix for all possible image pair combinations ($i,\:j$) where
($i=1,2...m;\:j=1,2...m$) is as follows:
\begin{equation}\label{covdef}
cov(\alpha_i,\beta_j)\:=\:
\left(\begin{array}{cccccc}
cov(\alpha_1,\beta_1) & cov(\alpha_1,\beta_2) & \:\:\:. & \:\:\:. & \:\:\:. & cov(\alpha_1,\beta_m)\\
cov(\alpha_2,\beta_1) & cov(\alpha_2,\beta_2) & \:\:\:. & \:\:\:. & \:\:\:. & cov(\alpha_2,\beta_m)\\
         .                 &           .                & \:\:\:. & \:\:\:  & \:\:\:  &            .              \\
         .                 &           .                & \:\:\:  & \:\:\:. & \:\:\:  &            .              \\
         .                 &           .                & \:\:\:  & \:\:\:  & \:\:\:. &            .              \\
cov(\alpha_m,\beta_1) & cov(\alpha_m,\beta_2) & \:\:\:. & \:\:\:. & \:\:\:. & cov(\alpha_m,\beta_m)\\
\end{array}\right)
\end{equation}

Figure~\ref{fig11} shows greyscale representations of covariance matrices
for all combinations of offset parameters (Equation~\ref{allcovs})
for the 105-image IRAC test case assuming {\it both relative and absolute} astrometric
source matches in the refinement. With 105 images, each covariance matrix has
$105\times105$ elements (or pixels in this representation). It is important
to note that the cross-correlation between different offset types is not
symmetric. For instance, the correlation between $\theta_i$ and $X_j$
is not the same as $\theta_j$ and $X_i$. Figure~\ref{fig12} shows the same
set but with {\it only relative} (frame-to-frame)
matches used for the refinement.
Comparing Figures~\ref{fig11} and~\ref{fig12},
we draw the following conclusions: first, for the case including astrometric
matches, ``long-distance'' correlations in image offsets are greatly reduced.
This is due to astrometric sources anchoring each image to the
fiducial reference frame, making them more independent of each other. 
The covariance matrices become essentially block-diagonal.
Second, uncertainties (variances along the diagonal for the same offset pair
combination) are greatly attenuated when astrometric matches are used
compared to the relative-only match case. There are greater numbers of
degrees of freedom per image when astrometrics are included, and this
reduces the relative uncertainty.

The relative-only case exhibits greater long-distance correlations
(larger off-diagonal values) since image positions are dictated solely by
frame-to-frame matches in image overlap regions.
This makes each successive image position dependent on it's nearest
neighbor positions, which depend on their own neighbours and so on
throughout the system of linked images.
Apart from astrometric matches reducing long distance correlations, this can
also happen if prior image pointing uncertainties are intrinsically smaller.
Small priors will pull refinement offsets from the global minimization
towards zero, which minimizes the $L_{apriori}$ term in Equation~\ref{cost}.
The priors force a constraint on each individual image to prevent
``over-refinement'', regardless of the number of source matches present.
Each image therefore becomes more independent of its neighbors, and
long-distance correlations are reduced.

We also note the rich and diverse patterns in the covariance
matrices for this observation, especially
the relative-only case (Figure~\ref{fig12}).
These are characteristic of the image layout and mosaic map geometry
(see Figure~\ref{fig8}). For instance, take the $cov(Y_i,Y_j)$
matrix in Figure~\ref{fig12}. The checker-board pattern arises from the
relative image numbering in the map and how this translates to the
numbering of elements in the covariance matrix (Equation~\ref{covdef}).
In the mosaic, the images repeat from right to left to create a leg,
then left to right and down again,
for a total of seven legs. Adjacent image pairs ($i,\:j$) along vertical
sections in the map are strongly correlated in their
($Y_i,\:Y_j$) offsets, while widely separated images are less correlated.
For instance, image number one
at top right has its $Y$ offset strongly correlated with the $Y$ offsets
of images directly below it (i.e., image numbers 30, 31, 60, 61, 90 and 91).
The high correlations are therefore with every $\sim$30th
image giving bright regions in the greyscale covariance image.
Also, the $Y$ offset of image number one is least correlated with that of
images in the far left vertical strip (image numbers 15, 16, 45, 46, 75,
76 and 105), resulting in dark regions in the covariance image.

In the end, one purpose of the covariance matrix is to visualize the
degree of correlation between image positions in the ensemble as a whole.
This allows one to ascertain whether refinement solutions are driven
by any particular invalid input assumptions (e.g., priors), or 
insufficient astrometric reference source information if robust absolute
refinement is desired.

\notetoeditor{Please place Figure~\ref{fig11} here: fig11.eps}

\notetoeditor{Please place Figure~\ref{fig12} here: fig12.eps}

\clearpage

\begin{figure}
\figurenum{1}
\plotone{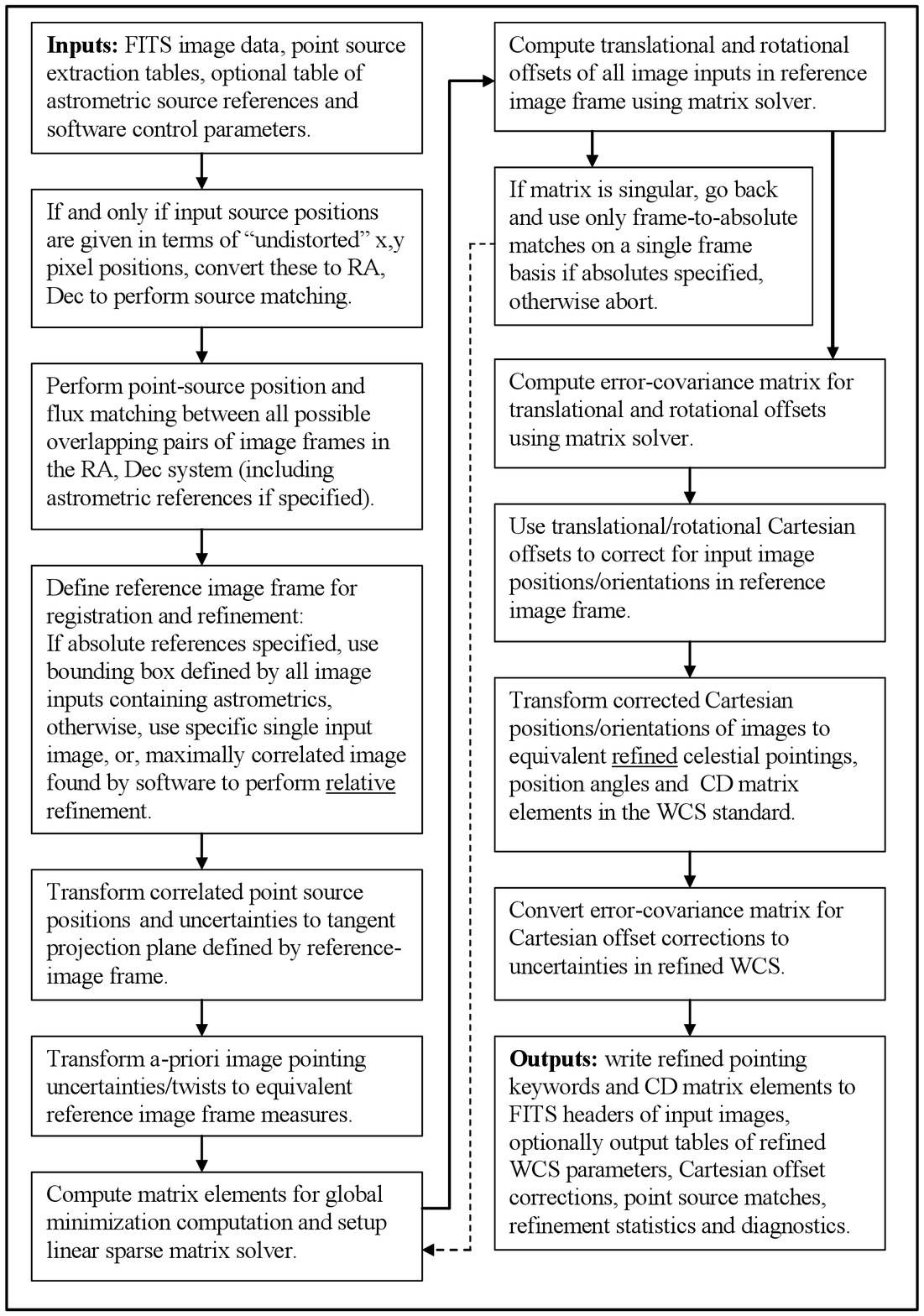}
\vspace{-40mm}
\caption{Processing and algorithmic flow in {\it pointingrefine}
software.
\label{fig1}}
\end{figure}

\clearpage 

\begin{figure}
\figurenum{2}
\plotone{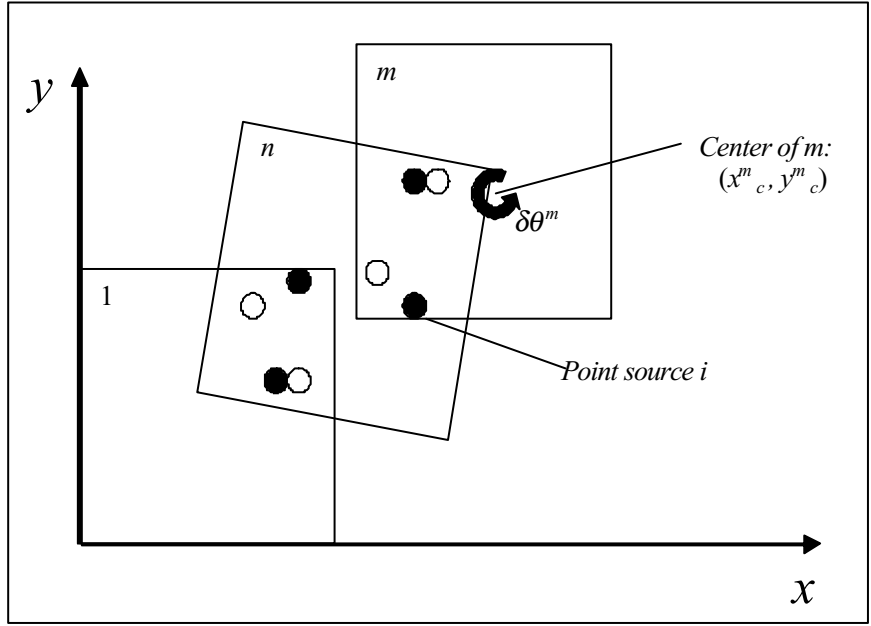}
\vspace{-110mm}
\caption{A simple three-image mosaic. Filled circles are sources detected in image
{\it n} and open circles are sources in {\it m} or 1.
\label{fig2}}
\end{figure}

\clearpage

\begin{figure}
\figurenum{3}
\plotone{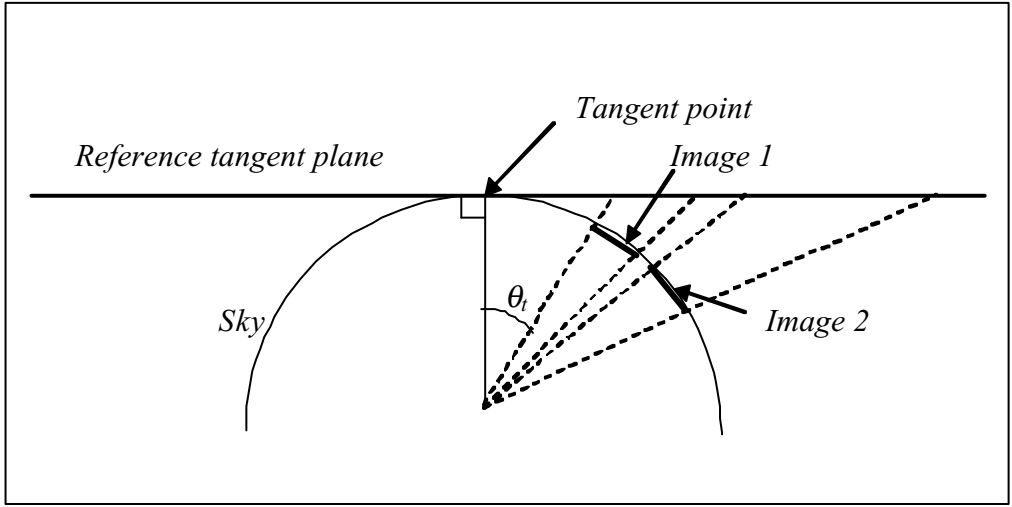}
\vspace{-100mm}
\caption{One-dimensional representation of projection geometry showing
images on sky and in tangent plane of reference image.
Images 1 and 2 have the same physical size but different projected
sizes.
\label{fig3}}
\end{figure}

\clearpage

\begin{figure}
\figurenum{4}
\epsscale{0.9}
\plottwo{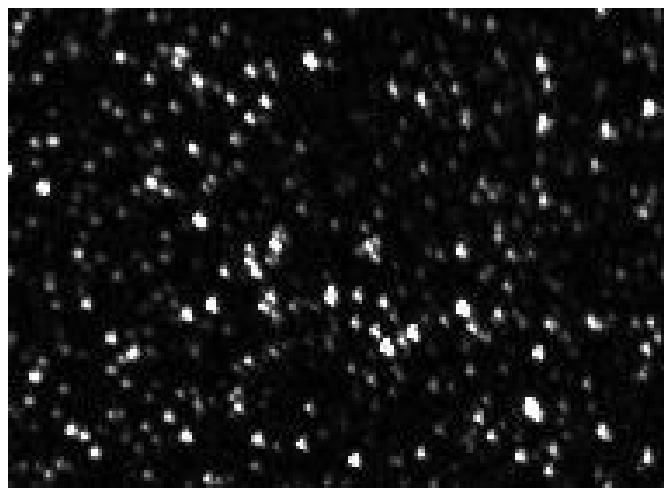}{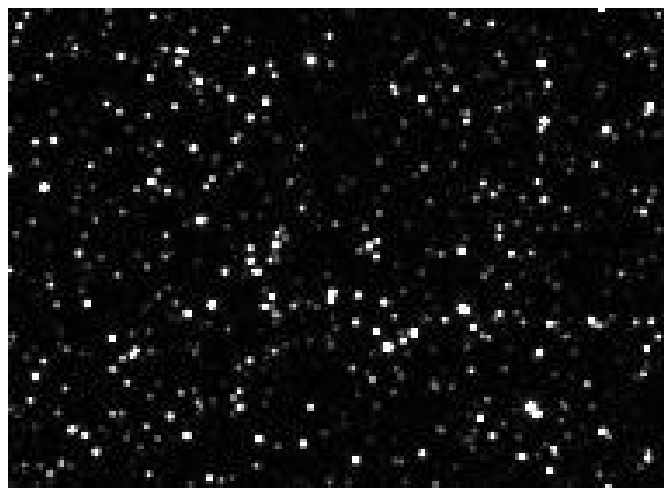}
\caption{{\it Unrefined} mosaic section of 1000-image IRAC simulation
at $3.6\micron$ on left and same section {\it after} pointing
refinement on right. Field dimensions are $\simeq2.8\arcmin\times3.8\arcmin$.
\label{fig4}}
\end{figure}

\clearpage

\begin{figure}
\figurenum{5}
\epsscale{1.4}
\plottwo{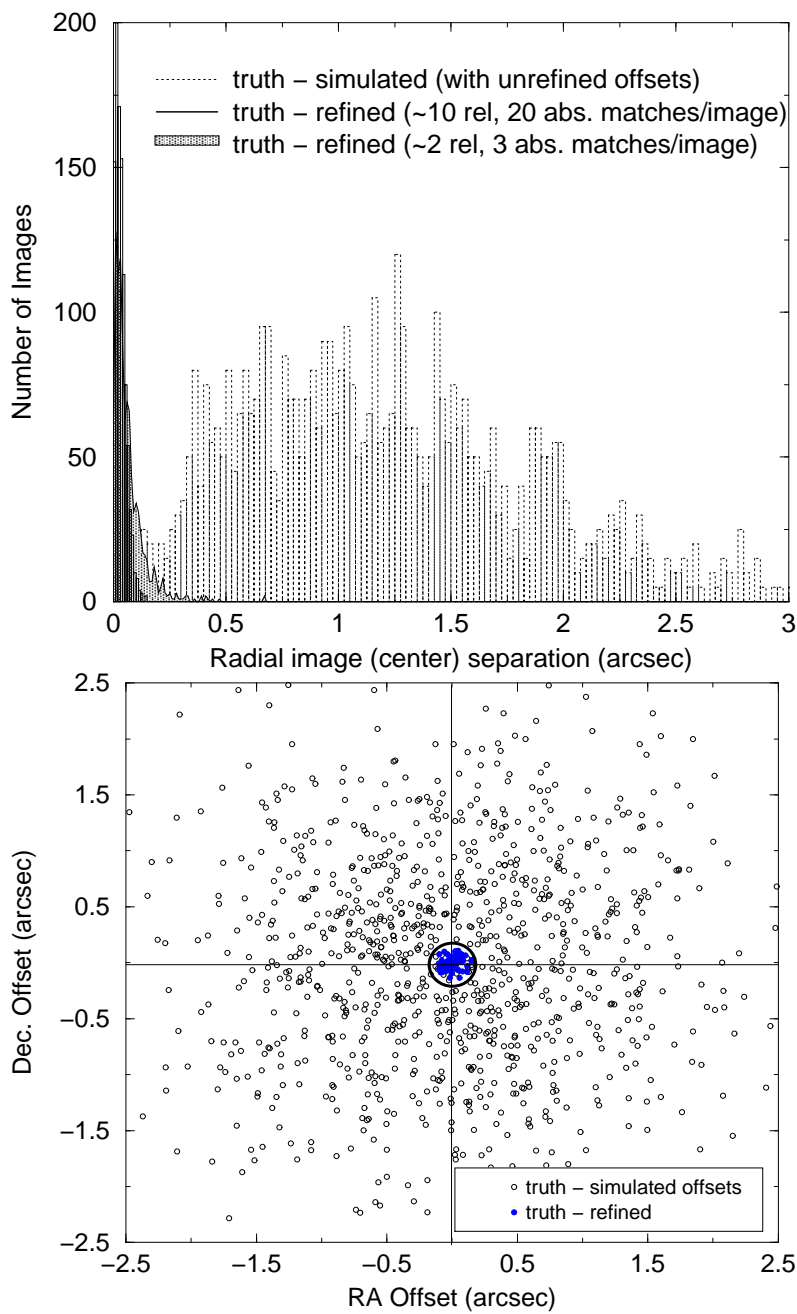}{fig5b.eps}
\caption{{\it Top}: distributions of image center separations relative to ``truth''
before and after refinement. {\it Bottom}: Offsets in RA and Dec between image centers
relative to truth. Circle represents 2-$\sigma$ region of
{\it truth - refined} distribution for case with $\sim 2$ relative/3 absolute matches
per image (dot-filled histogram in top figure).
\label{fig5}}
\end{figure}

\clearpage

\begin{figure}
\figurenum{6}
\epsscale{1.4}
\plottwo{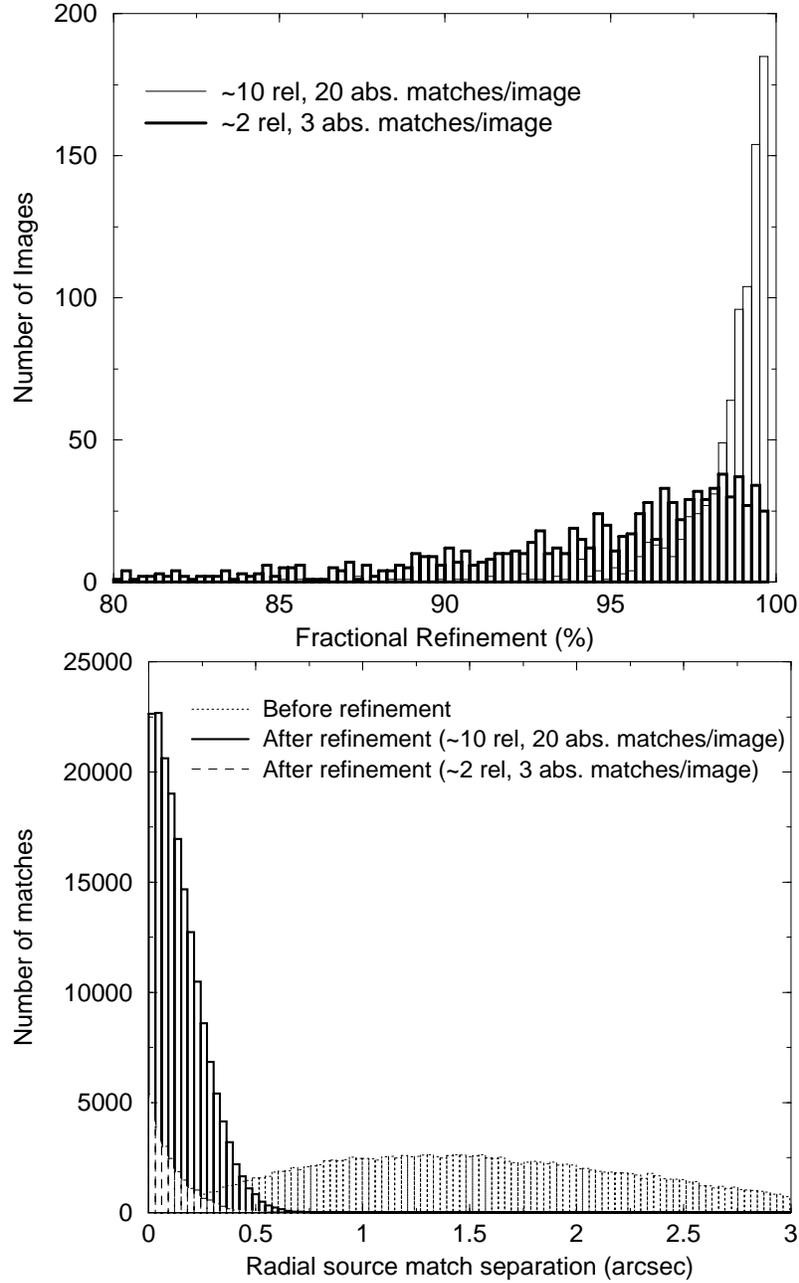}{fig6b.eps}
\caption{{\it Top}: distributions of the magnitude of refinement,
represented as a percentage of initial {\it truth - unrefined} image separation.
{\it Bottom}: distributions of matched source radial separations before and after
refinement. 
\label{fig6}}
\end{figure}

\clearpage

\begin{figure}
\figurenum{7}
\epsscale{1.4}
\plottwo{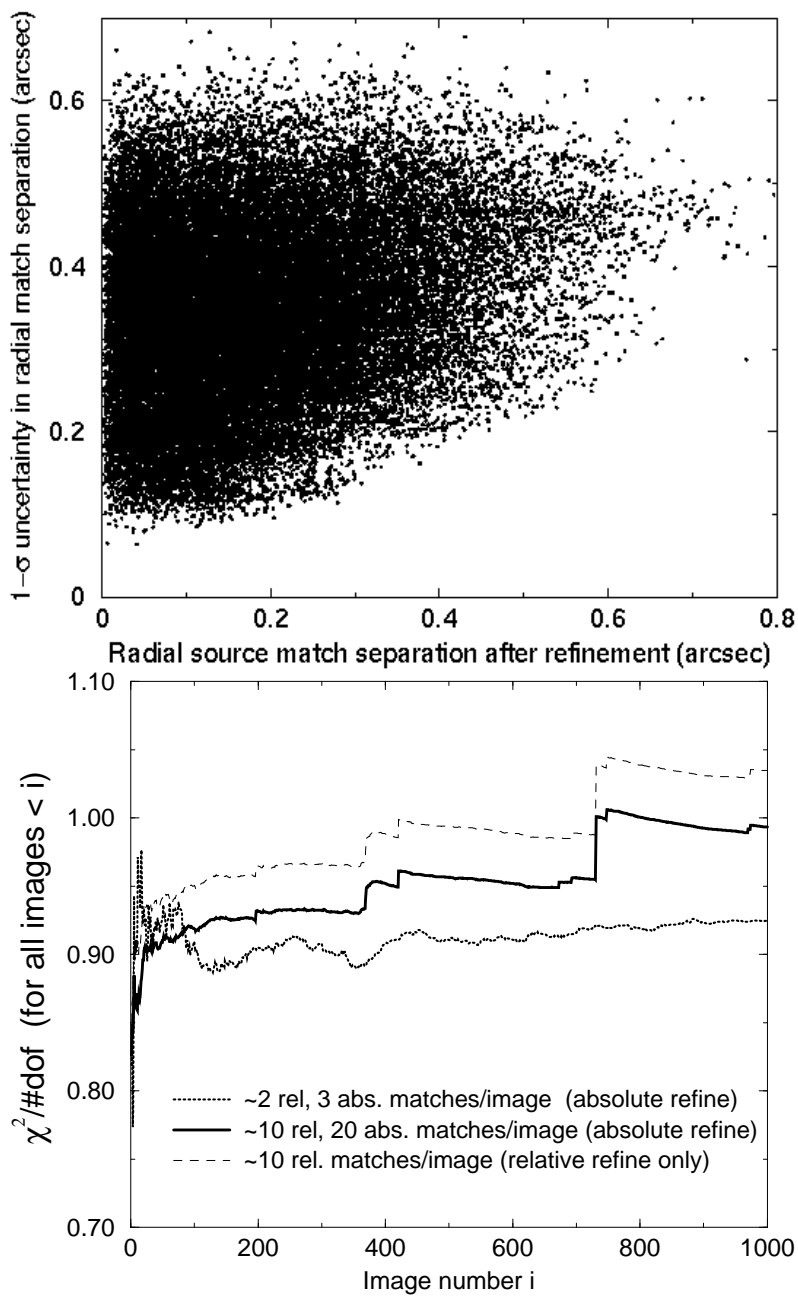}{fig7b.eps}
\caption{{\it Top}: uncertainty in matched source
radial separation (from centroid uncertainties) as a function of
actual source separation after refinement.
{\it Bottom}: reduced $\chi^2$ (i.e., $\chi^2$/number of degrees of
freedom) as a function of mosaic subset composed of successively
increasing numbers of images {\it i} (see Section~\ref{sim}).
\label{fig7}}
\end{figure}

\clearpage

\begin{figure}
\figurenum{8}
\epsscale{0.8}
\plotone{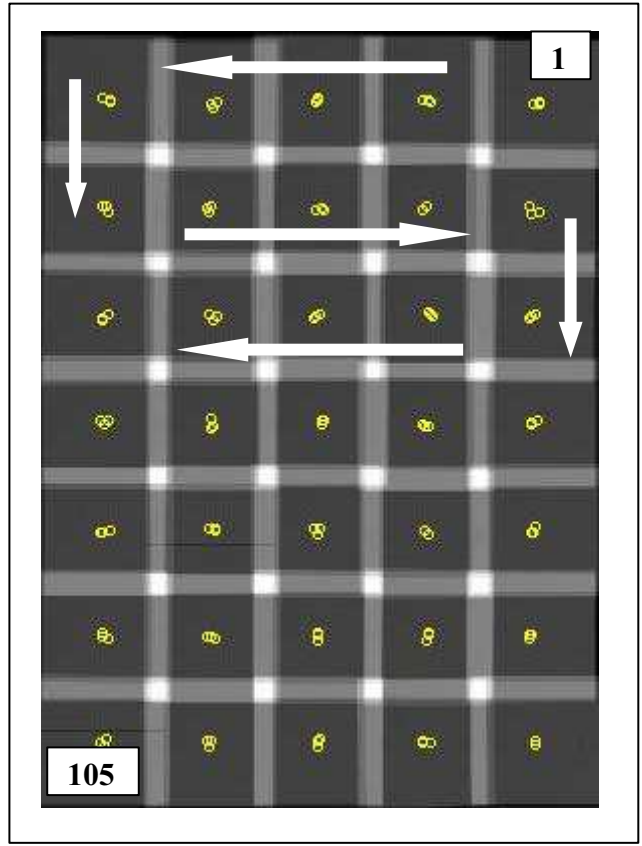}
\vspace{-130mm}
\caption{Coverage mosaic and mapping geometry of the 105-image IRAC observation
used in this analysis. Adjacent images have $\sim 20\%$ overlap with a coverage
of 6 and 12 pixels at edges and (inner) corners respectively. Mapping direction
is shown by arrows starting at top right and ending at bottom left.
Open circles are image centers.
\label{fig8}}
\end{figure}

\clearpage

\begin{figure}
\figurenum{9}
\epsscale{0.45}
\plotone{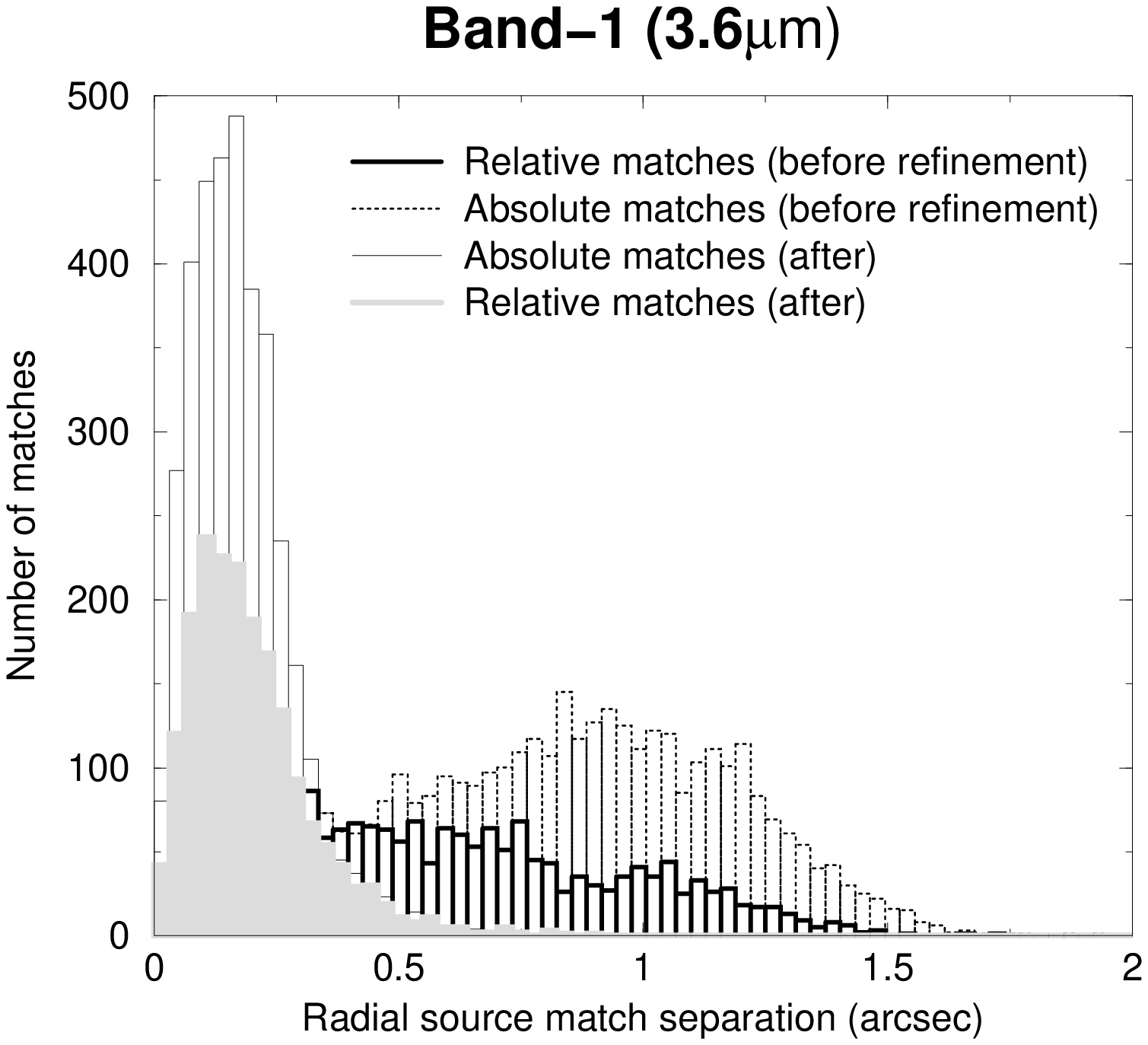}
\plotone{fig9b.eps}
\plotone{fig9c.eps}
\plotone{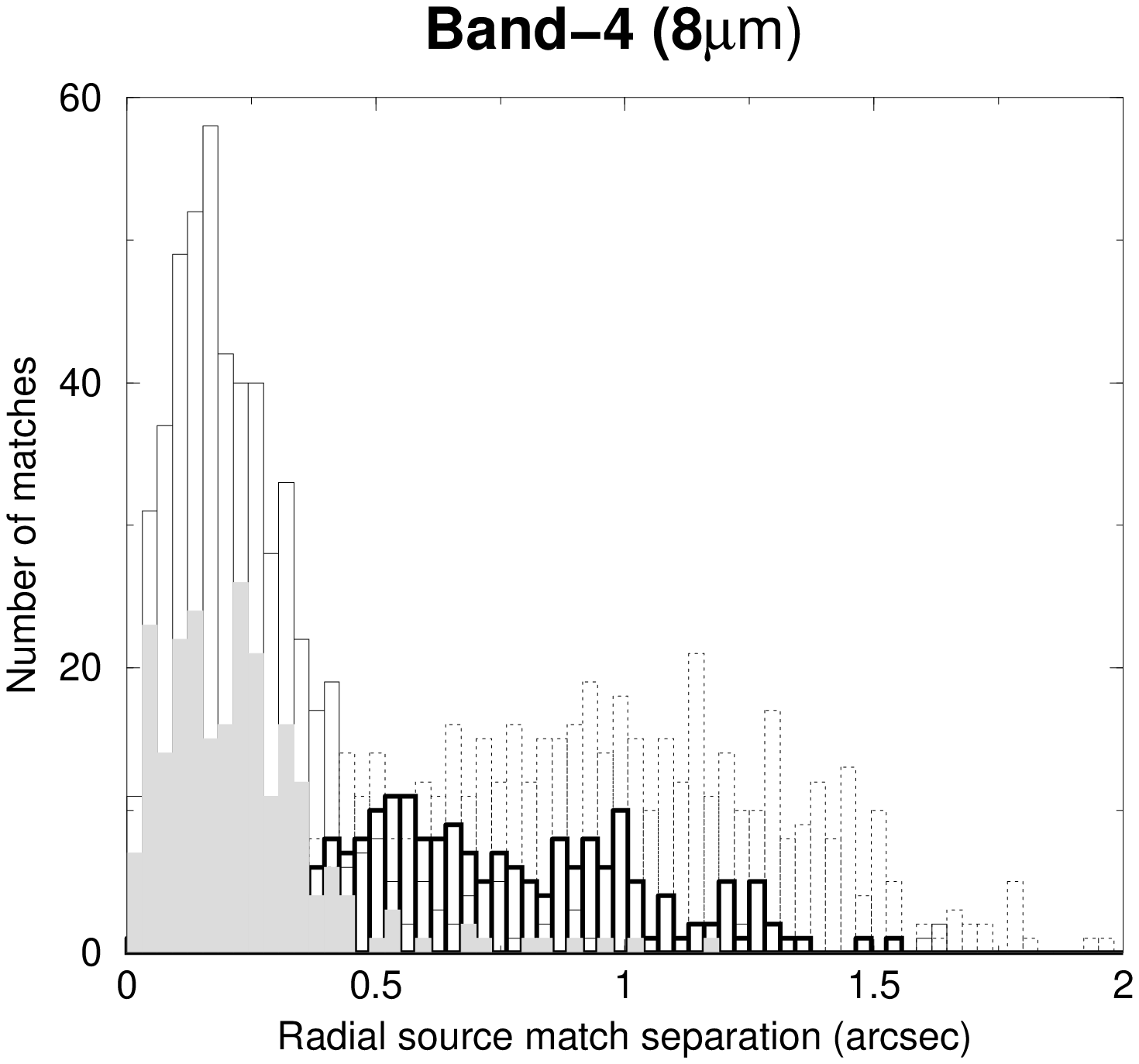}
\caption{Distributions of matched source radial separations before and after
refinement for all bands of {\it Spitzer's} IRAC instrument. Relative and
absolute astrometric (with {\it 2MASS} magnitudes $K_s\leq 15$)
matches have been separated.
\label{fig9}}
\end{figure}

\clearpage

\begin{figure}
\figurenum{10}
\epsscale{1.4}
\plottwo{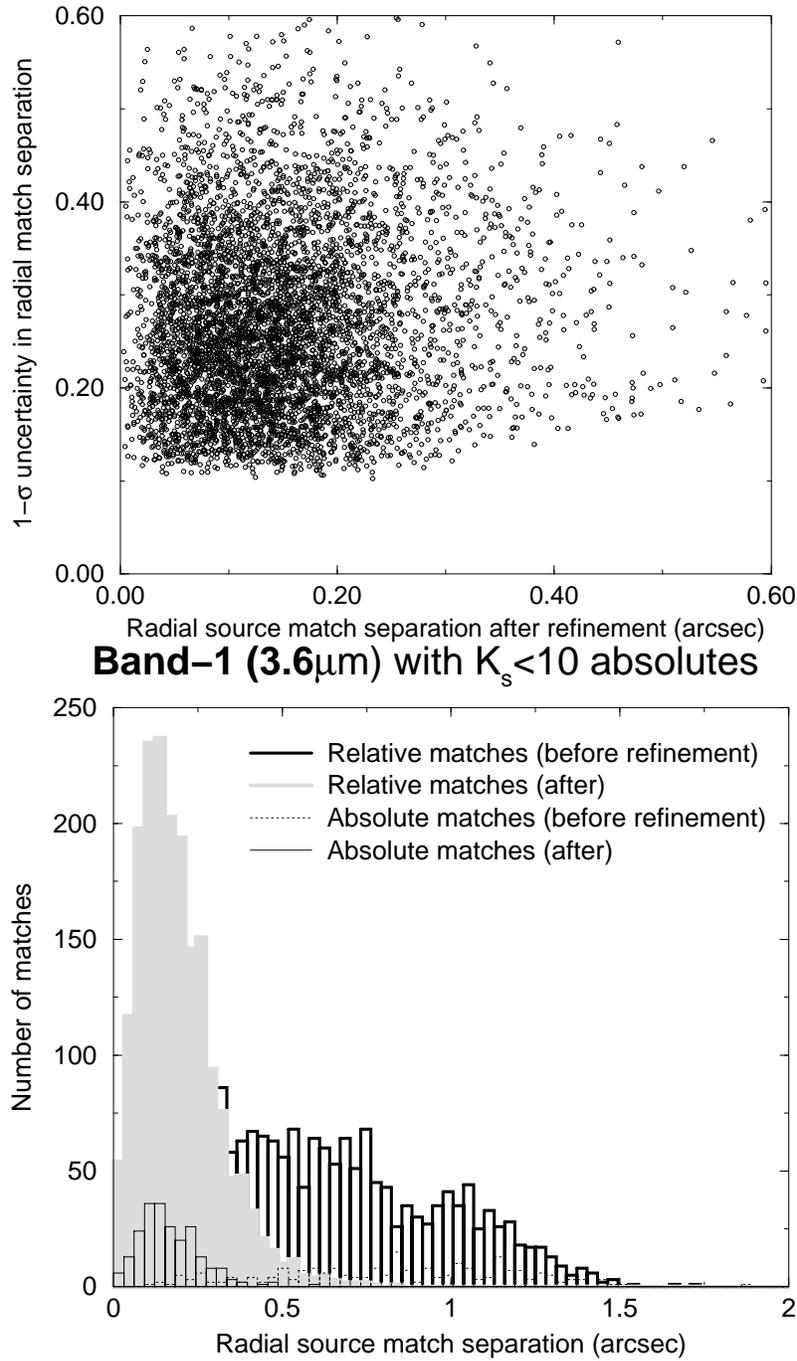}{fig10b.eps}
\caption{{\it Top}: uncertainty in matched source
radial separation (from centroid uncertainties of matches) as a function of
actual source separation after refinement for IRAC band-1.
{\it Bottom}: same as Figure~\ref{fig9} for IRAC band-1 but with
astrometric matches having a {\it 2MASS} magnitude limit of $K_s = 10$.   
\label{fig10}}
\end{figure}

\clearpage

\begin{figure}
\figurenum{11}
\epsscale{0.8}
\plotone{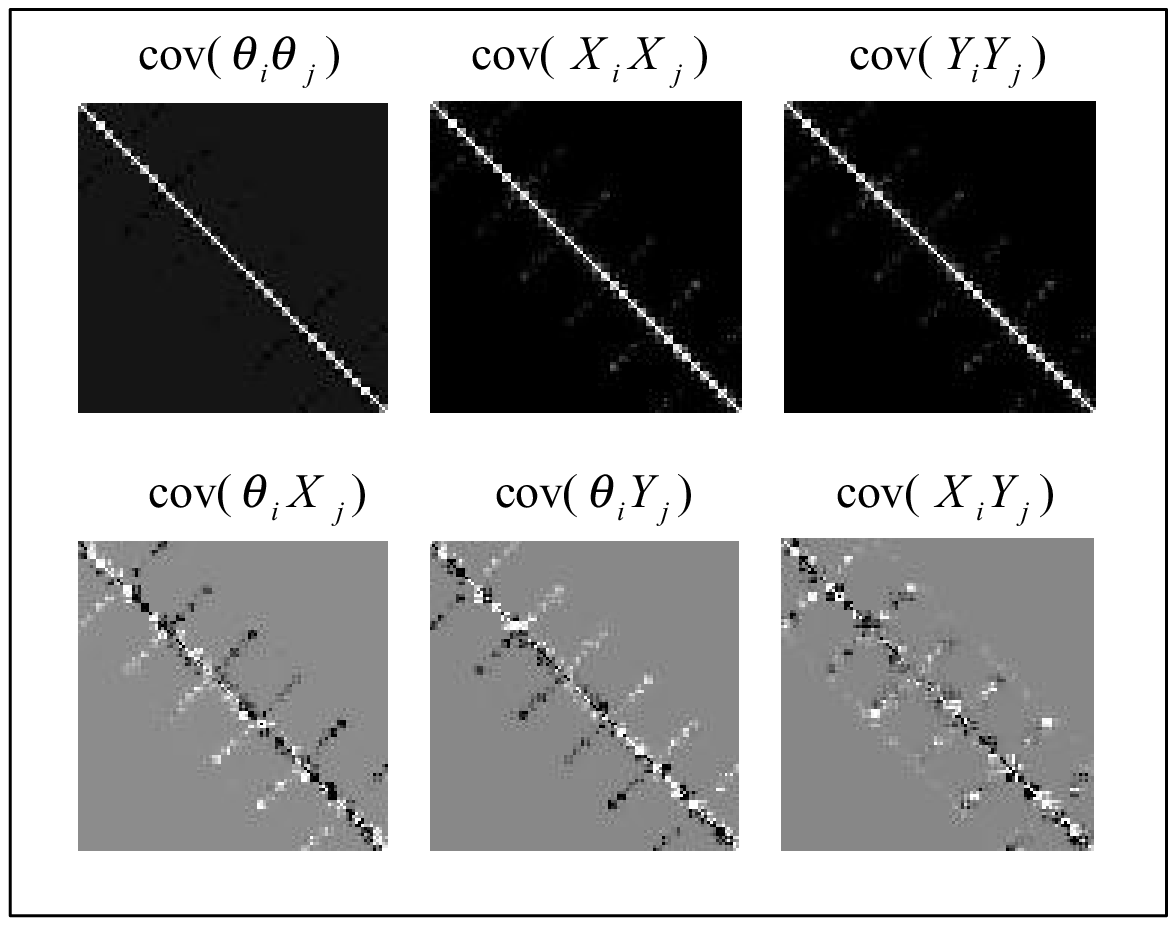}
\vspace{-100mm}
\caption{Greyscale representation of covariance matrices for
all orthogonal and rotational offsets in IRAC test case when
absolute astrometric matches are used. See Equation~\ref{covdef}
for matrix definition.   
\label{fig11}}
\end{figure}

\clearpage

\begin{figure}
\figurenum{12}
\epsscale{0.8}
\plotone{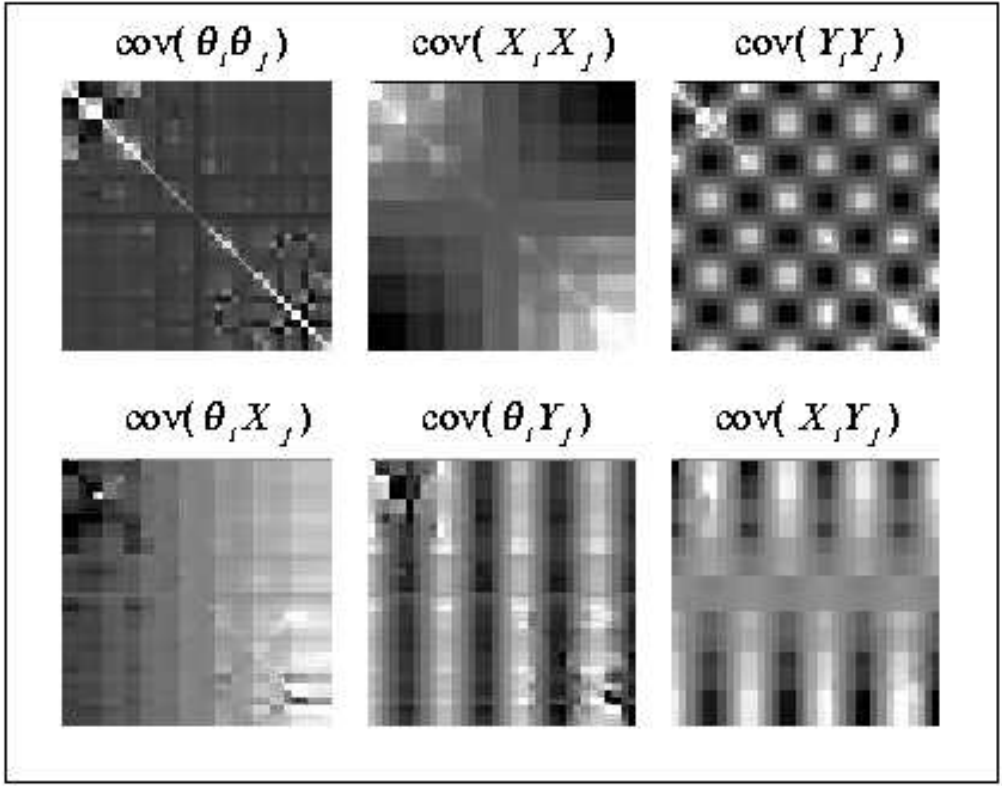}
\caption{Greyscale representation of covariance matrices for
all orthogonal and rotational offsets in IRAC test case when
absolute astrometric matches are {\underline {\it not}} used
(i.e. only relative frame-to-frame matches used).
See Equation~\ref{covdef} for matrix definition.   
\label{fig12}}
\end{figure}


\clearpage

\figcaption[fig1.eps]{Processing and algorithmic flow in
{\it pointingrefine} software.\label{fig1}}

\figcaption[fig2.eps]{A simple three-image mosaic. Filled circles are
sources detected from image {\it n} and open circles
are sources from {\it m} or 1.\label{fig2}}

\figcaption[fig3.eps]{One-dimensional representation of projection geometry showing
images on sky and in tangent plane of reference image.
Images 1 and 2 have the same physical size but different projectedsizes.\label{fig3}}

\figcaption[fig4a.eps, fig4b.eps]{{\it Unrefined}
mosaic section of 1000-image IRAC simulation
at $3.6\micron$ on left and same section {\it after} pointing
refinement on right. Field dimensions are
$\simeq2.8\arcmin\times3.8\arcmin$.\label{fig4}}

\figcaption[fig5a.eps, fig5b.eps]{{\it Top}: distributions of image center separations relative to ``truth''
before and after refinement. {\it Bottom}: Offsets in RA and Dec between image centers
relative to truth. Circle represents 2-$\sigma$ region of
{\it truth - refined} distribution for case with $\sim 2$ relative/3 absolute matches
per image (dot-filled histogram in top figure).\label{fig5}}

\figcaption[fig6a.eps, fig6b.eps]{{\it Top}: distributions of the magnitude of refinement,
represented as a percentage of initial {\it truth - unrefined} image separation.
{\it Bottom}: distributions of matched source radial separations before and after
refinement.\label{fig6}}

\figcaption[fig7a.eps, fig7b.eps]{
{\it Top}: uncertainty in matched source
radial separation (from centroid uncertainties) as a function of
actual source separation after refinement.
{\it Bottom}: reduced $\chi^2$ (i.e., $\chi^2$/number of degrees of
freedom) as a function of mosaic subset composed of successively
increasing numbers of images {\it i} (see Section~\ref{sim}).\label{fig7}}

\clearpage

\figcaption[fig8.eps]{Coverage mosaic and mapping geometry of the 105-image IRAC observation
used in this analysis. Adjacent images have $\sim 20\%$ overlap with a coverage
of 6 and 12 pixels at edges and (inner) corners respectively. Mapping direction
is shown by arrows starting at top right and ending at bottom left.
Open circles are image centers.\label{fig8}}

\figcaption[fig9a.eps, fig9b.eps, fig9c.eps, fig9d.eps]{
Distributions of 
matched source radial separations before and after
refinement for all bands of {\it Spitzer's} IRAC instrument. Relative and
absolute astrometric (with {\it 2MASS} magnitudes $K_s\leq 15$) matches have
been separated.\label{fig9}}

\figcaption[fig10a.eps, fig10b.eps]{
{\it Top}: uncertainty in matched source
radial separation (from centroid uncertainties of matches) as a function of
actual source separation after refinement for IRAC band-1.
{\it Bottom}: same as Figure~\ref{fig9} for IRAC band-1 but with
astrometric matches having a {\it 2MASS} magnitude limit of $K_s = 10$.   
\label{fig10}}

\figcaption[fig11.eps]{Greyscale representation of covariance matrices for
all orthogonal and rotational offsets in IRAC test case when
absolute astrometric matches are used. See Equation~\ref{covdef}
for matrix definition.\label{fig11}}

\figcaption[fig12.eps]{Greyscale representation of covariance matrices for
all orthogonal and rotational offsets in IRAC test case when
absolute astrometric matches are {\underline {\it not}} used
(i.e. only relative frame-to-frame matches used).
See Equation~\ref{covdef} for matrix definition.\label{fig12}}

\clearpage

\begin{deluxetable}{crrrrrrrc}
\footnotesize
\tablecaption{Statistics for IRAC-observation case study.\label{tab1}}
\tablewidth{0pt}
\tablehead{
\colhead{$\lambda$-Band} &
\colhead{$\langle \#Abs.\rangle$\tablenotemark{a}} &
\colhead{$\langle \#Rel.\rangle$\tablenotemark{b}} &
\colhead{$\langle D_{bef}\rangle_{A}$\tablenotemark{c}} &
\colhead{$\langle D_{bef}\rangle_{R}$\tablenotemark{c}} &
\colhead{$\langle D_{aft}\rangle_{T}$\tablenotemark{d}} &
\colhead{$\chi^2$} &
\colhead{dof\tablenotemark{e}} &
\colhead{$\chi^2/$dof}\nl
\colhead{\micron} &
\colhead{} &
\colhead{} &
\colhead{arcsec} &
\colhead{arcsec} &
\colhead{arcsec}
}

\startdata
$3.6$ &34.7 &20.5 &0.847 &0.605 &0.151 &8580.84 &10821 &0.792\nl
$4.5$ &20.7 &13.7 &0.883 &0.622 &0.152 &3843.93 &5645  &0.681\nl
$5.8$ &9.4  &7.6  &0.933 &0.603 &0.142 &1250.31 &2443  &0.511\nl
$8.0$ &5.3  &2.1  &0.913 &0.651 &0.203 &1240.50 &1249  &0.993\nl
\enddata

\tablenotetext{a}{Average number of absolute matches per frame.}
\tablenotetext{b}{Average number of relative (frame-to-frame) matches per frame.}
\tablenotetext{c}{$\langle D_{bef}\rangle_{A}$ and $\langle D_{bef}\rangle_{R}$
represent the mean matched source separation {\it before} refinement
for absolute and relative matches respectively (see Figure~\ref{fig9}).}
\tablenotetext{d}{$\langle D_{aft}\rangle_{T}$
represents the mean total matched source separation (for {\it both} relative
and abolute matches) {\it after} refinement.}
\tablenotetext{e}{Number of degrees of freedom (see Equation~\ref{dof}).}
\end{deluxetable}

\end{document}